\documentclass[11pt,prb,superscriptaddress,a4paper]{article}

\usepackage{amssymb,amsmath,amsthm}
\usepackage{texdraw}
\usepackage{epic,eepic}
\usepackage{graphicx}
\usepackage[all]{xy}
\newtheorem{thm}{Theorem}[section]
\newtheorem{prop}[thm]{Proposition}
\newtheorem{lem}{Lemma}

\newtheorem*{unnumrem}{Remark}
\theoremstyle{remark}
\newtheorem{rem}[thm]{Remark}

\newcounter{oftheorem}[section]
\newenvironment{mytheorem}[1]%
{\begin{trivlist}
     \renewcommand{\theoftheorem}{#1~\thesection.\arabic{oftheorem}}
     \refstepcounter{oftheorem}
     \item[\hspace{\labelsep}\bf\thesection.\arabic{oftheorem} #1.]}%
{\end{trivlist}}

\begin{document}

\title{Poisson Yang-Baxter maps with binomial Lax matrices}
%
%      {We Don't Have a Title Yet}
%

\author{
\textbf{Theodoros E. Kouloukas},\\
Department of Mathematics,\\
         University of Patras,
Patras, GR-265 00\\ Greece\\
\textbf{Vassilios G. Papageorgiou},\\
Department of Mathematics,\\
         University of Patras, GR-265 00
Patras,\\ Greece}

%\title[Left Kan Extensions Preserving Finite Products]
 %     {Left Kan Extensions Preserving Finite Products}
%\author{Panagis Karazeris}
%\address{Department of Mathematics, University of Patras,
%        Patras, Greece}
%\email{pkarazer@math.upatras.gr}
%\author{Grigoris Protsonis}
%\address{Department of Mathematics, University of Patras,
 %       Patras, Greece}
%\email{protsonis@master.math.upatras.gr}

%\thanks{The second author acknowledges the support
 %       of the Greek Scholarships Foundation}

%\thanks{
%\keywords{}

%\subjclass{}

\maketitle

\begin{abstract}
A construction of
multidimensional parametric Yang-Baxter maps is presented. The
corresponding Lax matrices are the symplectic leaves of first degree
matrix polynomials equipped with the Sklyanin bracket. These maps
are symplectic with respect to the reduced symplectic structure on
these leaves and provide examples of integrable mappings. An
interesting family of quadrirational symplectic YB maps on
$\mathbb{C}^4 \times \mathbb{C}^4$ with $3\times 3$ Lax matrices is
also presented.
\end{abstract}

%\pacs{02.30lk, 45.20Jj, 02.20Uw}
%\keywords{Yang Baxter equation, symplectic leaves, Sklyanin bracket, integrability}
%\maketitle

\tableofcontents
\newpage

%%%%%%%%%%%%%%%%%%%%%%%%%%%%%%%%%%%%%%%%%%%%%%%%%%%%%%%%%%%%%%%%%%%%%%%%%%%%%%%%%%%%%%%
%%%%%%%%%%%%%%%%%%%%%%%%   IntroductionIntroduction                 %%%%%%%%%%%%%%%%%%
%%%%%%%%%%%%%%%%%%%%%%%%%%%%%%%%%%%%%%%%%%%%%%%%%%%%%%%%%%%%%%%%%%%%%%%%%%%%%%%%%%%%%%
\section{Introduction} \label{intro}

Set theoretical solutions of the quantum Yang-Baxter equation have
extensively been studied by many authors after the pioneer work of
Drinfeld \cite{drin}. Even before that, examples of such solutions appeared in
\cite{skly} by Sklyanin. Weinstein and Xu \cite{wein}
proposed a construction of such solutions using the dressing action
of Poisson Lie groups \cite{STS}. This was generalized later in
\cite{lu}, in order to construct solutions on any group that
acts on itself and the action satisfies a compatibility condition. The
algebraic aspects of the Yang-Baxter equation were developed by
Etingof, Schedler and Soloviev \cite{ESS}.

Veselov  \cite{ves2,ves3}
connected the set theoretical solutions of the quantum Yang-Baxter equations
with integrable mappings. More specifically, he proved that for such
a solution, that admits a Lax matrix, there is a hierarchy of
commuting transfer maps which preserve the spectrum of the
corresponding monodromy matrix. Furthermore he proposed the shorter term `Yang Baxter maps'
for the set theoretical solutions of the quantum Yang-Baxter
equation.

Yang-Baxter maps are closely related with
integrable equations on quad-graphs. This is due to the {\em
multidimensional consistency property} of these equations, 
introduced in \cite{BS,Ni}, which
in a way seems to be equivalent with the Yang-Baxter property.
An explicit classification of equations on quad-graphs with fields
in $\mathbb{C}$ that satisfy the 3-dimensional consistency property and
of the Yang-Baxter maps on $\mathbb{CP}^1 \times \mathbb{CP}^1$ is given in
\cite{ABS1} and \cite{ABS2} respectively (see also \cite{pstv}).
Higher dimensional Yang-Baxter maps are obtained from multi-field integrable lattice equations
through symmetry reduction \cite{skin,skin1}.

Loop groups equipped with the Sklyanin bracket
provide a natural framework in order to derive Yang-Baxter maps with polynomial Lax matrices.
In \cite{ves1} one of the most fundamental examples of a
parametric Yang-Baxter map, Adler's map, is given by Hamiltonian reduction of the loop group
$LGL_2(\mathbb{R})$. Based on these ideas, a construction of Poisson parametric Yang-Baxter maps
with first degree polynomial $2\times 2$ Lax matrices was presented by the authors \cite{kp} from
a re-factorization procedure guided by the conservation of the Casimir functions
under the maps. By considering a
complete set of Casimir functions, symplectic multi–parametric Yang-Baxter maps were derived
with explicit formulae in terms of matrix operations.

The purpose of this work is to generalize the method of \cite{kp} in order to derive
symplectic Yang-Baxter maps with Lax matrices that are obtained by reduction on symplectic leaves of
binomial matrices.

The necessary definitions and notation about YB maps and Lax
matrices, are given in section \ref{sec1}.
Section \ref{main} contains the main theory of the
construction of symplectic Yang-Baxter maps associated to $2\times2$ Lax matrices.
This is generalized in higher dimensions in section
\ref{ndim} using further assumptions. A general re-factorization formula of $n \times n$
binomial matrices is presented. A reduction procedure of $3 \times 3$ binomial matrices to four dimensional
symplectic leaves, provides a family of quadrirational,
symplectic YB maps on $\mathbb{C}^4 \times \mathbb{C}^4$. Finally we conclude in section \ref{persp} by giving
some comments and perspectives for future work.

%%%%%%%%%%%%%%%%%%%%%%%%%%%%%%%%%%%%%%%%%%%%%%%%%%%%%%%%%%%%%%%%%%%%%%%
%%%%%%%%%%%%%%%%%%%%%%%%%%%%%%%%%%%%%%%%%%%%%%%%%%%%%%%%%%%%%%%%%%%%%%%
%%%%%%%%%%%%%%%%%%%%%%%%%%%%%%%%%%%%%%%%%%%%%%%%%%%%%%%%%%%%%%%%%%%%%%%

\section{Yang-Baxter Maps and lax matrices} \label{sec1}

Let $\mathcal{X}$ be any set. A map $R: \mathcal{X} \times
\mathcal{X} \rightarrow \mathcal{X} \times \mathcal{X}$,
$R:(x,y)\mapsto (u(x,y),v(x,y))$, that satisfies the {\em Yang-Baxter
equation} :
\begin{equation}
R_{23}R_{13}R_{12}=R_{12}R_{13}R_{23} \label{YBprop}
\end{equation}
is called {\em Yang-Baxter Map (YB)} \cite{ves2}. Here by $R_{ij}$
for $i,j=1,...,3$, we denote the map that acts as $R$ on the $i$ and
$j$ factor of $\mathcal{X} \times \mathcal{X} \times \mathcal{X}$
and identically on the others i.e.
\begin{eqnarray*}
R_{12}(x,y,z)&=&(u(x,y),v(x,y),z), \\
R_{13}(x,y,z)&=&(u(x,z),y,v(x,z)), \\
R_{23}(x,y,z)&=&(x,u(y,z),v(y,z)),
\end{eqnarray*}
for $x, \ y, \ z \in \mathcal{X}$. From our point of view, we consider that the set $\mathcal{X}$ 
has the structure of an algebraic 
variety. The YB map $R$ is called {\em
non-degenerate} if the maps
$u(\cdot,y):\mathcal{X}\rightarrow\mathcal{X}$ and
$v(x,\cdot):\mathcal{X}\rightarrow\mathcal{X}$ are bijective maps
and {\em quadrirational} \cite{ABS2} if they are rational bijective
maps.

%%%%%%%%%%%%%%%%%%%%%%%%%%%%%%%%%%%%%%%%%%%%%%%%%%%%%%%%%%%%%%%%
Parametric YB maps appear in the study of integrable equations on
quad-graphs. A {\em parametric YB map} is a YB map:
\begin{equation} \label{pYB}
R:((x,\alpha),(y,\beta))\mapsto((u,\alpha),(v,\beta))=
((u(x,\alpha, y,\beta),\alpha),(v(x,\alpha, y,\beta),\beta))
\end{equation}
where $x, \ y \in \mathcal{X}$ and the parameters $\alpha, \beta \in
\mathbb{C}^n$. We usually keep the parameters separately and denote
$R(x,\alpha,y,\beta)$ by $R_{\alpha,\beta}(x,y)$. According to
\cite{ves4} a {\em Lax Matrix} for the YB map (\ref{pYB}) is
a matrix $L(x,\alpha, \zeta )$ that depends on the point $x$, the
parameter $\alpha$ and a spectral parameter $\zeta$ (we usually
denote it just by $L(x;\alpha)$), such that
\begin{equation} \label{laxmat}
L(u;\alpha)L(v;\beta)=L(y;\beta)L(x;\alpha),
\end{equation}
for any $\zeta\in \mathbb{C}$. Furthermore if equation
(\ref{laxmat}) is equivalent to $(u,\ v)=R_{\alpha,\beta}(x,y)$ then
we will call $L(x;\alpha)$ {\em strong Lax matrix}.

\begin{figure}[h]
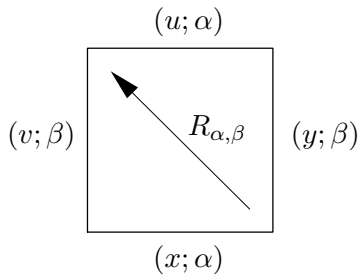

\centertexdraw{ \setunitscale 0.12 \linewd 0.01 \move (7 1) \linewd
0.03 \arrowheadtype t:F \avec(1 7) \lpatt( ) \move (0 0) \linewd
0.05 \lvec(8 0)  \lvec (8 8)  \lvec (0 8)  \lvec (0 0) \lpatt( )
\htext (2.8 -1.9){$(x;\alpha)$} \htext (8.75 3.5){$(y;\beta)$}
\htext (2.8 8.5){$(u;\alpha)$} \htext (-3.5 3.5){$(v;\beta)$} \htext
(4.3 3.8){$R_{\alpha,\beta}$} } \caption{A map assigned to the edges
of a quadrilateral} \label{fig:YBmap}
\end{figure}

A parametric YB map can be represented as a map assigned to the
edges of an elementary quadrilateral like in Fig.\ref{fig:YBmap}.

\begin{figure}[h]
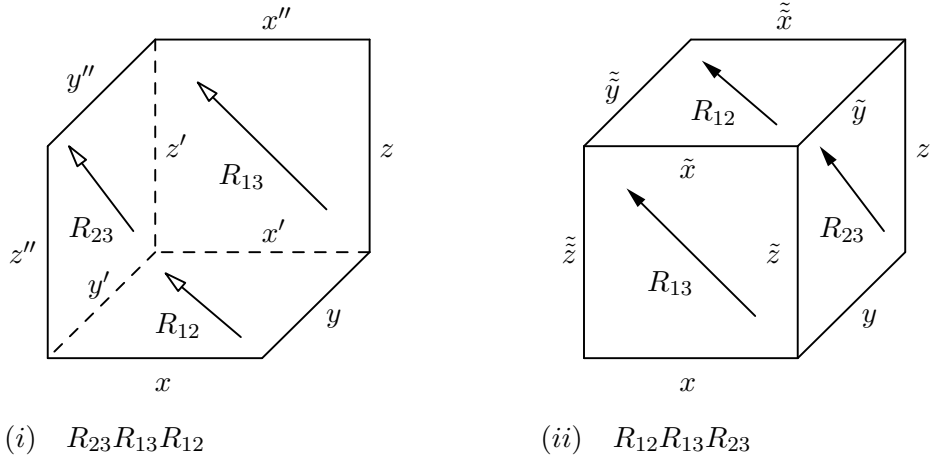

\centertexdraw{ \setunitscale 1.11 \linewd 0.06

\arrowheadsize l:0.1 w:0.05

\move (1. 0)  \linewd 0.01 \lvec(0.5 -0.5) \lpatt() \lvec(0.5
0.5)\lpatt() \lvec(1.0 1.0) \lpatt() \move(0.5 0.5)
\lpatt()\lvec(-0.5  0.5)\lpatt() \lvec(0.0 1.0)\lpatt(0.05 0.05)

\move(-2.5 0.0) \lpatt(0.05 0.05)
\lvec(-3.0 -0.5) \lpatt()
\lvec(-2.0 -0.5) \lpatt()
\lvec(-1.5 0.0) \lpatt(0.05 0.05)
\lvec(-2.5 -0.0)  \lpatt()
\move(-1.5 0.0) \lvec(-1.5 1.0)\lpatt() \lvec(-2.5 1.0)
\lpatt(0.05 0.05) \lvec(-2.5 0.0) \lpatt()
\move(-3.0 -0.5) \lvec(-3.0 0.5) \lpatt()

\lvec(-2.5 1.0) \lpatt()
\move(0 1) \lvec(1 1) \lpatt()\lvec(1 0)\lpatt()
\move(-0.5 0.5) \lvec(-0.5 -0.5) \lpatt() \lvec(0.5 -0.5)\lpatt()

\move(-2.6 0.1) \avec(-2.9 0.5)\lpatt()
\move(-1.7 0.2) \avec(-2.3 0.8)\lpatt()
\move(-2.1 -0.4) \avec(-2.45 -0.1)\lpatt()
\arrowheadtype t:F
\move(0.9 0.1) \avec(0.6 0.5)\lpatt()
\move(0.3 -0.3) \avec(-0.3 0.3)\lpatt()
\move(0.4 0.6) \avec(0.05 0.9)\lpatt()

\htext (-0.05 -0.65) {$x$}
\htext (0.8 -0.35) {$y$}
\htext (1.05 0.45) {$z$}
\htext (0.35 -0.05) {$\tilde{z}$}
\htext (0.75 0.6) {$\tilde{y}$}
\htext (-0.05 0.35) {$\tilde{x}$}
\htext (-0.6 -0.05) {$\tilde{\tilde{z}}$}
\htext (-0.4 0.7) {$\tilde{\tilde{y}}$}
\htext (0.4 1.05) {$\tilde{\tilde{x}}$}

\htext (-2.5 -0.65) {$x$}
\htext (-1.7 -0.35) {$y$}
\htext (-1.45 0.45) {$z$}
\htext (-2.0 0.05) {$x'$}
\htext (-2.81 -0.22) {$y'$}
\htext (-2.0 1.05) {$x''$}
\htext (-2.45 0.45) {$z'$}
\htext (-2.91 0.75) {$y''$}
\htext (-3.18 -0.05) {$z''$}

\htext (0.6 0.05) {$R_{23}$}
\htext (-0.2 -0.2) {$R_{13}$}
\htext (0.0 0.6) {$R_{12}$}

\htext (-2.5 -0.4) {$R_{12}$}
\htext (-2.2 0.3) {$R_{13}$}
\htext (-2.9 0.05) {$R_{23}$}

\htext (-3.2 -0.95)  {$(i) \ \ \ R_{23}R_{13}R_{12}$}
\htext (-0.7 -0.95) {$(ii) \ \ \ R_{12}R_{13}R_{23}$}

}
\caption{Cubic representation of the Yang--Baxter property}
\label{fig:CYB}
\end{figure}

We can also represent the maps $R_{23}R_{13}R_{12}$ and
$R_{12}R_{13}R_{23}$ as chains of maps at the faces of a cube like
in Fig.\ref{fig:CYB}. The first map corresponds to the composition
of the down, back, left faces, while the second one to the right,
front and upper faces. All the parallel edges to the $x$ (resp.
$y,z$) axis carry the parameter $\alpha$ (resp. $\beta$, $\gamma$).
If we denote by $(x^{\prime \prime },y^{\prime \prime },z^{\prime
\prime })$ and by $(\tilde{\tilde{x}},
\tilde{\tilde{y}},\tilde{\tilde{z}})$ the corresponding values
$R_{23}R_{13}R_{12}(x,y,z)$ and $R_{12}R_{13}R_{23}(x,y,z)$, then
Eq.(\ref{YBprop}) assures that $x''= \tilde{\tilde{x}},  \
y''=\tilde{\tilde{y}}$ and $z''=\tilde{\tilde{z}}$.
\\

The following proposition \cite{ves2,kp} gives a sufficient condition for
a solution of the Lax equation (\ref{laxmat}), in order to satisfy
the Yang-Baxter property.

\begin{prop} \label{anagea}
Let $u=u_{\alpha,\beta}(x,y)$, $v=v_{\alpha,\beta}(x,y) $ and
$A(x;\alpha)$ a matrix depending on a point $x$, a parameter
$\alpha$ and a spectral parameter $\zeta$, such that
$A(u;\alpha)A(v;\beta)=A(y;\beta)A(x;\alpha)$. If the equation
\begin{equation}
A( \hat{x}; \alpha )A( \hat{y}; \beta )A(\hat{z}; \gamma )=
A(x; \alpha )A(y; \beta )A(z; \gamma ) \label{xyz}
\end{equation}
implies that $\hat{x}=x, \ \hat{y}=y$ and $\hat{z}=z$,
then the map
$R_{\alpha,\beta}(x,y)=(u,v)$
is a parametric Yang-Baxter map with Lax matrix $A(x;\alpha)$.
\end{prop}

In a more general setting concerning integrable lattices
(not necessary YB maps), instead of the notion of a Lax matrix, the
notion of a {\em Lax pair} is more suitable. A Lax pair for a map $
\Phi_{\alpha,\beta}:((x,\alpha),(y,\beta))\mapsto((u,\alpha),(v,\beta))=
((u(x,\alpha, y,\beta),\alpha),(v(x,\alpha, y,\beta),\beta)) $ is a pair of matrices
$L, \ M$ depending on a point in $\mathcal{X}$, a parameter and a
spectral parameter $\zeta$ such that
\begin{equation} \label{laxpair}
L(u,\alpha,\zeta)M(v,\beta,\zeta)=M(y,\beta,\zeta)L(x,\alpha,\zeta),
\end{equation}
for any $\zeta\in \mathbb{C}$.
Combinations of Lax pairs can provide solutions of the entwining Yang-Baxter equation \cite{kp2}.

The dynamical aspects of the Yang-Baxter maps have been extensively
investigated in \cite{ves2} and \cite{ves3} where
commuting transfer maps, that preserve the spectrum of the
corresponding monodromy matrices, are introduced for each YB map.
These maps are believed to be integrable in the Liouville  sense,
i.e. symplectic mappings $M^{2n}\rightarrow M^{2n}$ that admit $n$
functionally independent integrals in involution.

%%%%%%%%%%%%%%%%%%%%%%%%%%%%%
%%%%%%%%%%%%%%%%%%%%%%%%%%%%%%%%%%%%%%%%%%%%%%%%%%%%%%%%%%%%%%%%%%%%%%%%%%%%%%%%%%%%%%%%%%%%%%
%%%%%%%%%%%%%%%%%%%%%%%%%%%%%%%%%%%%%%%%%%%%%%%%%%%%%%%%%%%%%%%%%%%%%%%%%%%%%%%%%%%%%%%%%%%%%%
%%%%%%%%%%%%%%%%%%%%%%%%%%%%%%%%%%%%%%%%%%%%%%%%%%%%%%%%%%%%%%%%%%%%%%%%%%%%%%%%%%%%%%%%%%%%%%

\section{Symplectic Yang--Baxter maps associated to binomial $2 \times 2$ Lax matrices} \label{main}
A general matrix re-factorization procedure provides a way of
constructing rational multi-parametric Yang-Baxter maps on
$\mathbb{C}^4\times \mathbb{C}^4$ with $2\times 2$ Lax matrices in
the form of first-degree matrix polynomials. These maps are Poisson
with respect to the Sklyanin bracket. By reduction on symplectic
leaves we derive 4-dimensional symplectic parametric YB maps. The
whole procedure generalizes the one presented in \cite{kp},
where the leading terms of the matrix polynomials were assumed
equal.

\subsection{Poisson Yang--Baxter maps from matrix re-factorization  } \label{sec2}
We consider the set
$\mathcal{L}^2$ of $2 \times 2$ polynomial matrices of the form $L(\zeta)=X-\zeta A$,
$\zeta\in\mathbb{C}$ equipped with the Sklyanin bracket \cite{skly2}:
\begin{equation}
\{L(\zeta ) \ \overset{\otimes }{,} \ L(\eta)\}=[\frac{r}{\zeta
-\eta},L(\zeta )\otimes L(\eta)],  \label{sklyanin}
\end{equation}
where $r$ denotes the permutation matrix: $r(x\otimes y) = y\otimes
x$. For
$$X=
\begin{pmatrix}
x _{1} & x_2 \\
x_3 & x_{4}
\end{pmatrix} \ {\text{and} } \ \
A=
\begin{pmatrix}
a_1 & a_2 \\
a_3 & a_4
\end{pmatrix},
$$
the brackets between the coordinate functions are given by the antisymmetric
Poisson structure matrix :
\begin{equation}
J_{A}(X)=
\begin{pmatrix}
0 & -x_{2}a_{1}+x_{1}a_{2} & x_{3}a_{1}-x_{1}a_{3} &
x_{3}a_{2}-x_{2}a_{3} \\
* & 0 & x_{4}a_{1}-x_{1}a_{4} &
x_{4}a_{2}-x_{2}a_{4} \\
* & * & 0 &
-x_{4}a_{3}+x_{3}a_{4} \\
* & * & *& 0
\end{pmatrix}
\label{strmatrix}
\end{equation}
where $J_{A}(X)_{ij}=\{ x_{i}- \zeta
a_{i},x_{j}- \zeta a_{j} \}$, for $i,j=1,...,4$.

There are six linear independent Casimir functions of
$\mathcal{L}^2$ which are the elements $a_i$, $i=1,...,4$, of the
matrix $A$ and the functions:
$$f_{0}(X;A)= \det X, \ \ f_{1}(X;A)=a_4x_1-a_3x_2-a_2x_3+a_1x_4,$$
i.e. the coefficients of the polynomial
$$p_{X}^A(\zeta):=\det(X-\zeta A)=
f_{2}(X;A) \zeta^{2}-f_{1}(X;A) \zeta + f_{0}(X;A)$$ with
$f_{2}(X;A)= \det A$ (of course $f_{2}(X;A)$ is also Casimir). For
any constant matrix $A$ we denote by $i_A$ the immersion $i_A: X
\mapsto X-\zeta A$ and by $\mathcal{L}^2_{A}$ the level set
$$\mathcal{L}^2_{A}=\{X-\zeta A \ | \ X \in Mat( 2\times 2) \}.$$
Furthermore for any pair of matrices $A,~B \in GL_2(\mathbb{C})$, we
define the matrix functions $\Pi_{A,B}^1, \ \Pi_{A,B}^2,$
with
\begin{eqnarray} \label{p1p2}
\Pi^{1} _{A,B}(X,Y)&=& f_{2}(X;A)(YA+BX)-f_{1}(X;A)AB, \\
\Pi^{2} _{A,B}(X,Y)&=& f_{2}(X;A)YX-f_{0}(X;A)AB .
\end{eqnarray}

\begin{prop}(re-factorization) \label{UV}
Let $A, \ B$ be invertible $2 \times 2$ matrices, such that $AB=BA$
and $X, Y \in Mat( 2\times 2)$ with $\det \Pi^{1} _{A,B}(X,Y) \neq 0$ .
Then
\begin{equation}
(U-\zeta A)(V-\zeta B)= (Y- \zeta B)(X-\zeta A), \label{fact}
\end{equation}
and $p_{U}^{A}(\zeta)=p_{X}^{A}(\zeta)$ (equivalently
$p_{V}^{B}(\zeta)=p_{Y}^{B}(\zeta)$), iff
\begin{eqnarray}  \label{U}
U=U_{A,B}(X,Y) &:=& \Pi^{2} _{A,B}(X,Y) \Pi^{1} _{A,B}(X,Y)^{-1}A, \\
V=V_{A,B}(X,Y) &:=& A^{-1}(YA+BX-U(X,Y)B). \label{V}
\end{eqnarray}
\end{prop}
The proof of this proposition is given in \cite{kp2}.

\begin{lem} \label{lemref}
Let $A_i, \ i=1,2,3$ be three invertible matrices such that
$A_iA_j=A_jA_i$, for $i,j=1,2,3$. Then
\begin{equation} \label{XYZ}
(X'_1-\zeta A_1)(X'_2-\zeta  A_2)(X'_3-\zeta  A_3)=(X_1-\zeta
A_1)(X_2-\zeta  A_2)(X_3-\zeta  A_3)
\end{equation}
and $p_{X'_i}^{A_{i}}(\zeta)=p_{X_i}^{A_{i}}(\zeta)$ for every $X_i
\in  Mat( 2\times 2), \ i=1,2,3$ and $\zeta \in \mathbb{C}$, iff
$X'_1=X_1$, $X'_2=X_2$ êáé $X'_3=X_3$.
\end{lem}
The proof of this lemma can be traced in the appendix of
\cite{kp2}.

\begin{prop} \label{gen2}
Let $K:\mathbb{C}^d \rightarrow GL_{2}(\mathbb{C})$, 
%%%%%%%%%%%%%%%%%%$V \subset \mathbb{C}^4$,
be a d--parametric family of commuting matrices.
%%%%%%%%%%%%%%a function such that $K(\alpha)K(\beta)=K(\beta)K(\alpha)$ for
For every $\alpha,\beta \in \mathbb{C}^d$ the map
\begin{eqnarray} \label{geYB}
\mathcal{R}_{\alpha,\beta}(X,Y)=(U_{K(\alpha),K(\beta)}(X,Y),V_{K(\alpha),K(\beta)}(X,Y)):=(U,V)
\end{eqnarray}
defined by (\ref{U}), (\ref{V}), is a parametric Yang-Baxter map
with Lax matrix $L(X;\alpha)=i_{K(\alpha)}(X)$ such that
$p_{U}^{K(\alpha)}(\zeta)=p_{X}^{K(\alpha)}(\zeta)$ and
$p_{V}^{K(\beta)}(\zeta)=p_{Y}^{K(\beta)}(\zeta)$.
\end{prop}

\paragraph{Proof:}
For $U=U_{K(\alpha),K(\beta)}(X,Y)$, $V=V_{K(\alpha),K(\beta)}(X,Y)$
and $L(X;\alpha)=i_{K(\alpha)}(X)$, from proposition \ref{UV} we
have that
$$L(U;\alpha)L(V;\beta)=L(Y;\beta)L(X;\alpha)$$
and $p_{U}^{K(\alpha)}(\zeta)=p_{X}^{K(\alpha)}(\zeta)$,
$p_{V}^{K(\beta)}(\zeta)=p_{Y}^{K(\beta)}(\zeta)$. Now, if we set
\begin{eqnarray*}
R_{\alpha,\beta}^{12}(X,Y,Z)&=&(X',Y',Z), \\
R_{\alpha,\gamma}^{13}\circ R_{\alpha,\beta}^{12}(X,Y,Z)&=&(X'',Y',Z'), \\
R_{\beta,\gamma}^{23}\circ R_{\alpha,\gamma}^{13}\circ
R_{\alpha,\beta}^{12}(X,Y,Z)&=& (X'',Y'',Z''),
\end{eqnarray*}
then $L(Y;\beta)L(X;\alpha)=L(X';\alpha)L(Y';\beta)$, and
$p_{X'}^{K(\alpha)}(\zeta)=p_{X}^{K(\alpha)}(\zeta)$,
$p_{Y'}^{K(\beta)}(\zeta)=p_{Y}^{K(\beta)}(\zeta)$. So
\begin{eqnarray*}
&&L(Z;\gamma)L(Y;\beta)L(X;\alpha)=(L(Z;\gamma)L(X';\alpha))L(Y';\beta)
 =L(X'';\alpha)(L(Z';\gamma)L(Y';\beta)) \\ && =
L(X'';\alpha)L(Y''\beta)L(Z'';\gamma) \
\end{eqnarray*}
$\text{and} \ p_{X''}^{K(\alpha)}(\zeta)=p_{X}^{K(\alpha)}(\zeta), \
p_{Y''}^{K(\beta)}(\zeta)=p_{Y}^{K(\beta)}(\zeta), \
p_{Z''}^{K(\gamma)}(\zeta)=p_{Z}^{K(\gamma)}(\zeta).$ \\
On the other hand for
\begin{eqnarray*}
 R_{\beta,\gamma}^{23}(X,Y,Z)&=&(X,\tilde{Y},\tilde{Z}), \\
R_{\alpha,\gamma}^{13}\circ R_{\beta,\gamma}^{23}(X,Y,Z)&=&(\tilde{X},\tilde{Y},\tilde{\tilde{Z}}), \\
R_{\alpha,\beta}^{12}\circ R_{\alpha,\gamma}^{13}\circ
R_{\beta,\gamma}^{23}(X,Y,Z)&=&
(\tilde{\tilde{X}},\tilde{\tilde{Y}},\tilde{\tilde{Z}})
\end{eqnarray*}
we get $L(Z;\gamma)L(Y;\beta)L(X;\alpha)=
L(\tilde{\tilde{X}};\alpha)L(\tilde{\tilde{Y}};\beta)L(\tilde{\tilde{Z}};\gamma)$
and
$p_{\tilde{\tilde{X}}}^{K(\alpha)}(\zeta)=p_{X}^{K(\alpha)}(\zeta)$,
$p_{\tilde{\tilde{Y}}}^{K(\beta)}(\zeta)=p_{Y}^{K(\beta)}(\zeta)$,
$p_{\tilde{\tilde{Z}}}^{K(\gamma)}(\zeta)=p_{Z}^{K(\gamma)}(\zeta).$
So finally we have that
\begin{eqnarray*}
 L(X'';\alpha)L(Y''\beta)L(Z'';\gamma)&=&
L(\tilde{\tilde{X}};\alpha)L(\tilde{\tilde{Y}};\beta)L(\tilde{\tilde{Z}};\gamma),
\\
p_{X''}^{K(\alpha)}(\zeta)~=~
p_{\tilde{\tilde{X}}}^{K(\alpha)}(\zeta), \
p_{Y''}^{K(\beta)}(\zeta) &=&
p_{\tilde{\tilde{Y}}}^{K(\beta)}(\zeta), \
p_{Z''}^{K(\gamma)}(\zeta)~=~p_{\tilde{\tilde{Z}}}^{K(\gamma)}(\zeta)
\end{eqnarray*}
and from lemma \ref{lemref} we derive $X''= \tilde{\tilde{X}},  \
Y''=\tilde{\tilde{Y}}, \ Z''=\tilde{\tilde{Z}}$, i.e.
$$R_{\beta,\gamma}^{23}\circ R_{\alpha,\gamma}^{13}\circ
R_{\alpha,\beta}^{12}= R_{\alpha,\beta}^{12}\circ
R_{\alpha,\gamma}^{13}\circ R_{\beta,\gamma}^{23}.$$

We will refer to the Yang-Baxter map of Prop. \ref{gen2} as the {\em
general parametric Yang-Baxter map associated with the function
$K$}. We have to notice that in general the Lax matrix
$L(X;\alpha)=i_{K(\alpha)}(X)$ is not a strong Lax matrix. For example
by considering $K(\alpha)=B$ for a constant $B \in
GL_2(\mathbb{C})$, the equation $i_B(U)i_B(V)=i_B(Y)i_B(X)$ except
of the corresponding solution (\ref{U}),(\ref{V}), admits also the
trivial solution $U=Y, \ V=X$ (elementary involution).

Now we return to the Poisson structure (\ref{strmatrix}).
We can extend the Poisson bracket of $\mathcal{L}^2$ to the Cartesian product
$\mathcal{L}^2\times \mathcal{L}^2$ as follows :
\begin{equation}
\{x_{i},x_{j}\}=J_{A}(X)_{ij},\ \{y_{i},y_{j}\}=J_{B}(Y)_{ij},\
\{x_{i},y_{j}\}=0,  \label{bracket}
\end{equation}
for any $( X-\zeta A,\ Y-\zeta B )\in \mathcal{L}^2\times \mathcal{L}^2$ where
$x_{i},\ x_{j},\ y_{i},\ y_{j}$ for $i=1,...,4$ are the elements of the
matrices $X,\ Y$ respectively.

\begin{prop} \label{pois}
The map $ \mathcal{R}:\mathcal{L}^2_{K(\alpha)}\times
\mathcal{L}^2_{K(\beta)} \rightarrow \mathcal{L}^2_{K(\alpha)}\times
\mathcal{L}^2_{K(\beta)}$,
\begin{equation} \label{pmap}
\mathcal{R}:(X-\zeta K(\alpha),Y-\zeta K(\beta)) \mapsto
(U_{K(\alpha),K(\beta)}(X,Y)- \zeta
K(\alpha),V_{K(\alpha),K(\beta)}(X,Y)-\zeta K(\beta))
\end{equation}
 is a Poisson map.
\end{prop}
\paragraph{Proof:}
A direct computation of the Poisson brackets of the elements of
$U=U_{K(\alpha),K(\beta)}(X,Y)$ and $V=V_{K(\alpha),K(\beta)}(X,Y)$
defined by (\ref{U}), (\ref{V}) gives:
$$\{u_{i},u_{j}\}=J_{K_{\alpha}}(U)_{ij},\ \{v_{i},v_{j}\}=J_{K_{\beta}}(V)_{ij},\
\{u_{i},v_{j}\}=0,$$
for $i=1,...,4$.

If we consider the permutation map $r:(X,Y) \mapsto (Y,X)$ and the
multiplication map $m:(X,Y) \mapsto XY$, then $\mathcal{R}$ is the
unique map defined by the commutative diagram:

$$
\xymatrix{\ \mathcal{L}^2_{k(\alpha) } \times
\mathcal{L}^2_{k(\beta)} \ar[dd]_r \ar[rr]^{\mathcal{R}}
& & \mathcal{L}^2_{k(\alpha) }  \times \mathcal{L}^2_{k(\beta) }  \ar[dd]^{~ m} \\
&  \\
\mathcal{L}^2_{k(\beta)} \times \mathcal{L}^2_{k(\alpha)} \ar[rr]^m
&  & \mathcal{L}^2_{2}      }
$$
$$Commutative \ diagram$$
Here $\mathcal{L}^2_{2}$ denotes the second degree polynomial
$2\times 2$ matrices. From proposition \ref{pois} and the
multiplication property of the Sklyanin bracket we conclude that
each map of this diagram is Poisson.

%%%%%%%%%%%%%%%%%%%%%%%%%%%%%%%%%%%%%%%%%%%%%%%%%%%%%%%%%%%%%%%%%%%%%%%%%%%%%%%%%%%%%%%%%%%%%%%%%%%%%%%
%%%%%%%%%%%%%%%%%%%%%%%%%%%%%%%%%%%%%%%%%%%%%%%%%%%%%%%%%%%%%%%%%%%%%%%%%%%%%%%%%%%%%%%%%%%%%%%%%%%%%%%
%%%%%%%%%%%%%%%%%%%%%%%%%%%%%%%%%%%%%%%%%%%%%%%%%%%%%%%%%%%%%%%%%%%%%%%%%%%%%%%%%%%%%%%%%%%%%%%%%%%%%%%

%------------------------------------------------------------------------------------------------------
\subsection{Reduction on symplectic leaves} \label{section32}

In the previous section it was pointed out that the matrix $A$ of a
generic element
$$X-\zeta A=
\begin{pmatrix}
x _{1} & x_2 \\
x_3 & x_{4}
\end{pmatrix} -\zeta
\begin{pmatrix}
a_1 & a_2 \\
a_3 & a_4
\end{pmatrix} \in \mathcal{L}^2,$$
belongs to the center of the Sklyanin algebra. In the four
dimensional Poisson submanifold $\mathcal{L}^2_A$ there are two
Casimir functions
$$f_{0}(X;A)= \det X \ \text{êáé} \ \ f_{1}(X;A)=a_4x_1-a_3x_2-a_2x_3+a_1x_4.$$
We restrict on the level set of the Casimir functions by solving the
system $f_{0}(X;{A})=\alpha_{0}$, $f_{1}(X;{A})=\alpha_{1}$ with
respect to two elements $x_i,x_j$ of $X$. So we consider two
functions $h_A, \ g_A$, defined on an open set $D \subset
\mathbb{C}^4$, such that
\begin{equation} \label{hg}
x_i=h_Á(x_k,x_l,\alpha_0,\alpha_1) \ \text{and} \
x_j=g_A(x_k,x_l,\alpha_0,\alpha_1) , \ k,l \notin \{i,j\}.
\end{equation}
We denote by $pr_{k,l}$ the projection of a matrix to its $k,l$
elements (by ordering the elements of a matrix from one to four as
before) and by $Pr$ the map
$$Pr=pr_{k,l} \times pr_{k,l}:(X,Y) \mapsto
(pr_{k,l}(X),pr_{k,l}(Y)).$$
By substituting the $x_i,x_j$ to the matrix $X$ we define the
parametric matrix $L'_A(x_{k},x_{l};\alpha_0,\alpha_1)$. For
simplicity we renumber $x_k \mapsto x_1$, $x_l \mapsto x_2$ and we
come up to the matrix $L'_A(x_{1},x_{2};\alpha_0,\alpha_1)$ that
satisfies the following equations
\begin{equation*}
f_{0}(L'_A(x_{1},x_{2};\alpha_0,\alpha_1);A) = \alpha_{0} \
\text{êáé} \ f_{1}(L'_A(x_{1},x_{2};\alpha_0,\alpha_1);A) =
\alpha_{1}.
\end{equation*}
The connected components of
$\Sigma_{A}(\alpha_0,\alpha_1)=\{L'_A(x_{1},x_{2};\alpha_0,\alpha_1)
-\zeta A \, | \ x_1,x_2 \in D\subset \mathbb{C} \}$ are two
dimensional symplectic leaves of $\mathcal{L}^2_A$.

By the next proposition the general YB map
$\mathcal{R}_{\alpha,\beta}$ of Prop. \ref{gen2} is reduced on the symplectic
leaves $\Sigma_{K(\alpha)}(\alpha_0,\alpha_1) \times
\Sigma_{K(\beta)}(\beta_0,\beta_1)$ of $\mathcal{L}^2 \times
\mathcal{L}^2$.

\begin{prop} \label{sympYB}
Let $K:\mathbb{C}^d  \mapsto GL_{2}(\mathbb{C})$
be a d--parametric family of commuting matrices. 
For every $\alpha,\beta \in \mathbb{C}^d $, the map
\begin{equation} \label{symYB}
R _{\bar{\alpha},\bar{\beta}}((x_1,x_2),(y_1,y_2))=Pr\circ
\mathcal{R}_{\alpha,\beta} (L'_{K(\alpha)}(x_1,x_2
;\alpha_0,\alpha_1),L'_{K(\beta )}(y_1,y_2;\beta_0,\beta_1)),
\end{equation}
is a non-degenerate symplectic Yang-Baxter map with vector
parameters $\bar{\alpha}=(\alpha,\alpha_{0}, \alpha_{1})$,
$\bar{\beta}=(\beta,\beta_{0},\beta_{1})\in V \times \mathbb{C}^{2}$
and strong Lax matrix
\begin{equation} \label{symlax}
L(x_1,x_2;\bar{\alpha})=i_{K(\alpha)}(L'_{K(\alpha)}(x_1,x_2;\alpha_0,\alpha_1))=
L'_{K(\alpha)}(x_1,x_2;\alpha_0,\alpha_1)- \zeta K(\alpha).
\end{equation}
\end{prop}

\paragraph{Proof:}
For $X=L'_{K(\alpha)}(x_1,x_2;\alpha_0,\alpha_1)$ and
$Y=L'_{K(\beta)}(y_1,y_2;\beta_0,\beta_1)$ we define the matrices
$U=U_{K(\alpha),K(\beta)}(X,Y), \ V=V_{K(\alpha),K(\beta)}(X,Y))$ by
(\ref{U}), (\ref{V})
$$(U,V)=\mathcal{R}_{\alpha,\beta}(X,Y)=\mathcal{R}_{\alpha,\beta}(L'_{K(\alpha)}(x_{1},x_{2};\alpha,\alpha_{0} ,\alpha_{1} ),
L'_{K(\beta)}(y_{1},y_{2};\beta,\beta_{0} ,\beta_{1})).$$ Since
$f_{i}(U;K(\alpha))=f_{i}(X;K(\alpha))=\alpha_i$ and
$f_{i}(V;K(\beta))=f_{i}(Y;K(\beta))=\beta_i$ for $i=0,1$, then
$U=L'_{K(\alpha)}(u_{1},u_{2};\alpha_{0} ,\alpha_{1} )$ and
$V=L'_{K(\beta)}(v_{1},v_{2};\beta_{0} ,\beta_{1})$. The projection
$Pr(U,V)$ gives the corresponding elements $u=(u_{1},u_{2})$ and
$v=(v_{1},v_{2})$). So the YB property of the map
$R_{\bar{\alpha},\bar{\beta}}:((x_1,x_2;\bar{\alpha}),(y_1,y_2;\bar{\beta}))
\mapsto((u_1,u_2,\bar{\alpha}),(v_1,v_2,\bar{\beta}))$,
 is immediately derived from the YB property of the Poisson map
$\mathcal{R}_{\alpha,\beta}$. Furthermore proposition \ref{gen2}
implies that
$i_{K(\alpha)}(U)i_{K(\beta)}(V)=i_{K(\beta)}(Y)i_{K(\alpha)}(X)$,
so
\begin{eqnarray} \nonumber
&~&(L'_{K(\alpha)}(u_{1},u_{2};\alpha_{0} ,\alpha_{1} )-\zeta
K_{\alpha}) (L'_{K(\beta)}(v_{1},v_{2};\beta_{0} ,\beta_{1})-\zeta
K_{\beta})\\
&=& (L'_{K(\beta)}(y_{1},y_{2};\beta_{0},\beta_{1})-\zeta
K_{\beta})(L'_{K(\alpha)}(x_{1},x_{2};\alpha_{0} ,\alpha_{1} )-\zeta
K_{\alpha}) \label{laxpr}
\end{eqnarray}
 which means that
$L(x_{1},x_{2};\bar{\alpha})=L'_{K(\alpha)}(x_{1},x_{2};\alpha_{0},\alpha_{1}
)-\zeta K_{\alpha}$ is a Lax matrix for
$R_{\bar{\alpha},\bar{\beta}}$. Also, from proposition \ref{UV} we
conclude that $L(x_{1},x_{2};\bar{\alpha} )$ is a strong Lax matrix.
Finally we notice that equation (\ref{laxpr}) is directly solvable
with respect to $v=(v_1, v_2)$ and $x=( x_1,  x_2)$, since
\begin{eqnarray*}
K_{\beta}^{-1}L'_{K_{\beta}}(v;\hat{\beta} )
=
%{\small=}
(L'_{K_{\alpha}}(u;\hat{\alpha} )
K_{\beta}-L'_{K_{\beta}}(y;\hat{\beta}
)K_{\alpha})^{-1}L'_{K_{\beta}}(y;\hat{\beta} )
K_{\beta}^{-1}(L'_{K_{\alpha}}(u;\hat{\alpha} )K_{\beta}-L'_{K_{\beta}}(y;\hat{\beta} )K_{\alpha}), ~~ \\
K_{\alpha}^{-1}L'_{K_{\alpha}}(x;\hat{\alpha} )
=
%{\small=}
(L'_{K_{\beta}}(y;\hat{\beta} )
K_{\alpha}-L'_{K_{\alpha}}(u;\hat{\alpha}
)K_{\beta})^{-1}L'_{K_{\alpha}}(u;\hat{\alpha} )
K_{\alpha}^{-1}(L'_{K_{\beta}}(y;\hat{\beta}
)K_{\alpha}-L'_{K_{\alpha}}(u;\hat{\alpha} )K_{\beta}) ~~~
\end{eqnarray*}
for $y=(y_1,y_2)$, $u=(u_1,u_2)$, $\hat{\alpha}=(\alpha_0,\alpha_1)$
and $\hat{\beta}=(\beta_0,\beta_1)$. That proves the non-degeneracy
of the YB map (\ref{symYB}).

\begin{rem}
From the construction of the Lax matrix $
L(x_{1},x_{2};\bar{\alpha})$ and lemma \ref{lemref} we can prove
that the equation:
$$ L(x'_{1},x'_{2};\bar{\alpha})L(y'_{1},y'_{2};\bar{\beta})L(z'_{1},z'_{2};\bar{\gamma})
=L(x_{1},x_{2};\bar{\alpha})L(y_{1},y_{2};\bar{\beta})L(z_{1},z_{2};\bar{\gamma})$$
implies $x'=x, \ y'=y$ and $z'=z$ (without further assumptions). So the YB property of the map
(\ref{symYB}) can be derived directly from Prop. \ref{anagea}.

\end{rem}

\begin{rem} \label{rempar}
If we set $\alpha_{0}=\beta_{0}=k$ on the YB map  (\ref{symYB}) we
obtain the parametric YB map $ R_{ \bar{\alpha}, \bar{\beta} }$ with
parameters $\bar{\alpha}=(\alpha,a_1), \ \bar{\beta}=(\beta,b_1) \in
V \times \mathbb{C}$ and Lax matrix
$L(x_{1},x_{2};\alpha,\alpha_{1}):=L(x_{1},x_{2};\alpha,k,\alpha_{1}
)$. We have analogous results if we identify any other pair of
parameters. If we set $\bar{\alpha}=\bar{\beta}$ then we derive the
trivial solution $U=Y$, $V=X$, because this is the only
solution of Eq.(\ref{fact}) with $A=B$, $f_{0}(U;A)=f_{0}(Y;A)$
and $f_{1}(U;A)=f_{1}(Y;A)$.
\end{rem}

%%%%%%%%%%%%%%%%%%%%%%%%%%%%%%%%%%%%%%%%%%%%%%%%%%%%%%%%%%%%%%%%%%%%%%%%%%%%%%%%%%%%%%%%%
%%%%%%%%%%%%%%%%%%%%%%%%%%%%%%%%%%%%%%%%%%%%%%%%%%%%%%%%%%%%%%%%%%%%%%%%%%%%%%%%%%%%%%%%%
\subsection{Classification}
In this section we classify the quadrirational YB maps with $2
\times 2$ binomial Lax matrices of our construction. In
\cite{kp} a classification by Jordan normal forms was given
for the case $K(\alpha)=K(\beta)=B$, with $B$ a $2 \times 2$
constant matrix.  Here we give a more general classification in
order to include all the cases that we considered.
%%%%%%%%%%%%%%%%%%%%%%%%%%%%%%%%%%%%%%%%%%%%%%%%%%%%%%%%%%%%%%%%%%%%%
%%%%%%%%%%%%%%%%%%%%%%%%%%%%%%%%%%%%%%%%%%%%%%%%%%%%%%%%%%%%%%%%%%%%%%%
First we begin by determining the functions $K$ of proposition
\ref{gen2}. Actually we are going to consider the problem of families of commuting matrices
up to conjugation. One can bring one member of the family to its Jordan canonical form and find all matrices 
commuting with it. From this analysis we conclude that, up
to conjugation, there are only two (non-disjoint) families  of
commuting pairs of matrices
$$I) \ A=\begin{pmatrix}
a_{1} & 0 \\
0 & a_2 %
\end{pmatrix}, \
B=\begin{pmatrix}
b_{1} & 0 \\
0 & b_2 %
\end{pmatrix} \ \text{and} \ \
II) \ A=\begin{pmatrix}
a_{1} & a_2 \\
0 & a_1 %
\end{pmatrix}, \
B=\begin{pmatrix}
b_{1} & b_2 \\
0 & b_1 %
\end{pmatrix}.
$$
Since the equation (\ref{laxmat}) and the YB maps are invariant under conjugation we can
restrict to these two general cases of the function $K:\mathbb{C}^{2}\rightarrow GL_{2}(\mathbb{C})$.

The last step towards the classification is to examine the
relevance of the choice of variables in the construction of the Lax
matrix that we presented in the previous section. In the first case, where $K(\alpha)$ is a matrix  
of the first family for any $\alpha \in \mathbb{C}^{2}$, the equations 
\begin{equation} \label{eqCas}
f_{0}(X;K(\alpha))=\alpha_{0}, \ f_{1}(X;K(\alpha))=\alpha_{1}
\end{equation}
are solvable with respect to any pair $(x_i,x_j)$, for $i,j=1,...,4, \ i\neq j$, 
except of the pair $(x_2,x_3)$, while for a matrix $K(\alpha)$ of the second family the 
equations are solvable with respect to any pair $(x_i,x_j)$,$i,j=1,...,4, \ i\neq j$. 
Now, let
us suppose that, by solving  equations 
(\ref{eqCas}) in a different way,
we have derived two matrices
$L'_{K(\alpha)}(x_1,x_2;\alpha_0,\alpha_1)$,
$M'_{K(\alpha)}(x'_1,x'_2;\alpha_0,\alpha_1)$ such that
\begin{eqnarray*}
f_{0}(L'_{K(\alpha)}(x_1,x_2;\alpha_0,\alpha_1);K(\alpha)) &=&
\alpha_{0}, \
f_{1}(L'_{K(\alpha)}(x_1,x_2;\alpha_0,\alpha_1);K(\alpha) )=\alpha_{1} \ \text{and} \ \\
f_{0}(M'_{K(\alpha)}(x'_1,x'_2;\alpha_0,\alpha_1);K(\alpha)) &=&
\alpha_{0}, \
f_{1}(M'_{K(\alpha)}(x'_1,x'_2;\alpha_0,\alpha_1));K(\alpha))=\alpha_{1}.
\end{eqnarray*}
Then there is a local diffeomorphism $\phi_{\bar{\alpha}}:\mathbb{C}^2 \rightarrow \mathbb{C}^2$ 
($\bar{\alpha}=(\alpha,\alpha_0,\alpha_1) \in \mathbb{C}^4$), 
such that
$\phi_{\bar{\alpha}}:(x_1,x_2) \mapsto (x'_1,x'_2)$ and
$$ M'_{K(\alpha)}(\phi_{\bar{\alpha}}(x_1,x_1);\alpha_0,\alpha_1)=
L'_{K(\alpha)}(x_1,x_2;\alpha_0,\alpha_1).$$ Now if we denote by
$R_{\bar{\alpha},\bar{\beta}}, \ R'_{\bar{\alpha},\bar{\beta}}$ the
parametric YB maps with strong Lax matrices
$L(x_{1},x_{2};\bar{\alpha})=L'_{K(\alpha)}(x_1,x_2;\alpha_0,\alpha_1)-
\zeta K(\alpha)$ and
$M(x'_{1},x'_{2};\bar{\alpha})=M'_{K(\alpha)}(x'_1,x'_2;\alpha_0,\alpha_1)-
\zeta K(\alpha)$ respectively, then
\begin{equation} \label{eqYB}
(\phi_{\bar{\alpha}} \times \phi_{\bar{\beta}})\circ
R'_{\bar{\alpha},\bar{\beta}}=R_{\bar{\alpha},\bar{\beta}} \circ
(\phi_{\bar{\alpha}} \times \phi_{\bar{\beta}}).
\end{equation}

From the above analysis we conclude that every four parametric
non-degenerate YB map on $\mathbb{C}^2 \times \mathbb{C}^2$, of
proposition \ref{sympYB}, can be reduced up to equivalence
(\ref{eqYB}) and reparametrization (see also remark \ref{rempar})
into one of the following two cases.

\subsubsection*{Case I }

We consider the generic element $X- \zeta K_1(\alpha_1,\alpha_2) \in
\mathcal{L}^2_{K_1(\alpha_1,\alpha_2)}$ with
$$X=
\begin{pmatrix}
x _{1} & x_2 \\
x_3 & x_{4}
\end{pmatrix} \ \text{and} \
K_1(\alpha_1,\alpha_2)=\begin{pmatrix}
\alpha_{1} & 0 \\
0 & \alpha_2 %
\end{pmatrix}.$$
The Casimir functions in this case are
$$f_1(X;K_1(\alpha_1,\alpha_2))=\alpha_2
x_1 + \alpha_1 x_4, \ f_0(X;K_1(\alpha_1,\alpha_2))=x_1x_4-x_2x_3.$$
By setting $f_0(X;\alpha_1,\alpha_2)=\alpha_3, \
f_1(X;\alpha_1,\alpha_2)=\alpha_4$ and solving with respect to $x_3$,
$x_4$, for $\alpha_1,x_2 \neq 0$,  we derive the matrix
\begin{equation}
L'_{K_1(\hat{\alpha})}(x_1,x_2;\alpha_3,\alpha_4)= \left(
\begin{array}{cc}
 {x_1} & {x_2} \\
 \frac{{x_1} ({\alpha_4}-{\alpha_2} {x_1})-{\alpha_1}
   {\alpha_3}}{{\alpha_1} {x_2}} & \frac{{\alpha_4}-{\alpha_2}
   {x_1}}{{\alpha_1}}
\end{array}
\right) \ \text{with} \ \hat{\alpha}=(\alpha_1,\alpha_2)
\end{equation}
and the 8-parametric quadrirational YB map of proposition
\ref{sympYB}
\begin{equation*}
R^1 _{\bar{\alpha},\bar{\beta}}((x_1,x_2),(y_1,y_2))=Pr \circ
\mathcal{R}^1_{\hat{\alpha},\hat{\beta}}
(L'_{K_1(\hat{\alpha})}(x_1,x_2;\alpha_3,\alpha_4),L'_{K_1(\hat{\beta})}(y_1,y_2;\beta_3,\beta_4)).
\end{equation*}
Here $\mathcal{R}^1_{\hat{\alpha},\hat{\beta}}$ is the general
parametric YB map (\ref{geYB}) associated with the function $K_1$,
the projection $Pr= pr_{1,2} \times pr_{1,2}$ (projections at the
elements of the first arrow of a matrix) and the parameters are
$\bar{\alpha}=(\alpha_1,\alpha_2,\alpha_3,\alpha_4)$,
$\bar{\beta}=(\beta_1,\beta_2,\beta_3,\beta_4)$. According to prop.
\ref{sympYB}, this map admits the strong Lax matrix
$$L_1(x_1,x_2;\bar{\alpha})=L'_{K_1(\hat{\alpha})}(x_1,x_2;\alpha_3,\alpha_4)- \zeta
K_1(\alpha_1,\alpha_2),$$ and for $\alpha_1,\beta_1 \neq 0$ it is a symplectic
rational map  on $\{(x_1,x_2),(y_1,y_2) \in \mathbb{C}^2 \times
\mathbb{C}^2 \ | \ x_2,y_2 \neq 0\}$, with respect to the reduced
symplectic form defined by the brackets:
\begin{equation*}
\{x_1,x_2\}=-\alpha_1 x_2, \  \{y_1,y_2\}=-\beta_1 y_2, \
\{x_i,y_j\}=0 \ \text{for} \ i=1,2.
\end{equation*}

\subsubsection*{Case II }

For $K_2(\alpha_1,\alpha_2)=\begin{pmatrix}
\alpha_{1} & \alpha_{2} \\
0 & \alpha_1 %
\end{pmatrix}$ we set again $f_0(X;K_2(\alpha_1,\alpha_2))=\alpha_3, \
f_1(X;K_2(\alpha_1,\alpha_2))=\alpha_4$ and solve with respect to to $x_3$,
$x_4$ to get
\begin{equation}
L'_{K_2(\hat{\alpha})}(x_1,x_2;\alpha_3,\alpha_4)= \left(
\begin{array}{cc}
 {x_1} & {x_2} \\
 \frac{{\alpha_4} {x_1}-{\alpha_1}
   \left({x_1}^2+{\alpha_3}\right)}{{\alpha_1}
   {x_2}-{\alpha_2} {x_1}} & \frac{{\alpha_2}
   {\alpha_3}-{\alpha_4} {x_2}+{\alpha_1} {x_1}
   {x_2}}{{\alpha_2} {x_1}-{\alpha_1} {x_2}}
\end{array}
\right), \ \text{with} \ \hat{\alpha}=(\alpha_1,\alpha_2)
\end{equation}
and the corresponding YB map
\begin{eqnarray*}
R^2 _{\bar{\alpha},\bar{\beta}}((x_1,x_2),(y_1,y_2))=Pr \circ
\mathcal{R}^2_{\hat{\alpha},\hat{\beta}}
(L'_{K_2(\hat{\alpha})}(x_1,x_2;\alpha_3,\alpha_4),L'_{K_2(\hat{\beta})}(y_1,y_2;\beta_3,\beta_4)),
\end{eqnarray*}
with $ \bar{\alpha}=(\alpha_1,\alpha_2,\alpha_3,\alpha_4), \
\bar{\beta}=(\beta_1,\beta_2,\beta_3,\beta_4), \ Pr= pr_{1,2} \times
pr_{1,2}$ and $\mathcal{R}^2_{\hat{\alpha},\hat{\beta}}$ the general
parametric YB map associated with $K_2$. This map admits the strong
Lax matrix
$L_2(x_1,x_2;\bar{\alpha})=L'_{K_2(\hat{\alpha})}(x_1,x_2;\alpha_3,\alpha_4)-
\zeta K_2(\alpha_1,\alpha_2)$. The reduced Sklyanin bracket in this case is
given by brackets of the coordinates
\begin{equation*}
\{x_1,x_2\}=\alpha_2 x_1-\alpha_1 x_2, \  \{y_1,y_2\}=\beta_2 y_1-\beta_1 y_2, \
\{x_i,y_j\}=0 \ \text{for} \ i,j=1,2.
\end{equation*}

As it was pointed out, YB maps with less parameters can be
constructed from these two cases by setting $\alpha_i=\beta_i=k$ for
some $i \in \{1,2\}$. Also, by using appropriate scalings, one can reduce the number of parameters. However, we do not do this here, having in mind degenerate cases in subsection \ref{DGNYB} below,  as well as consideration of continuous limits in the future. 

\begin{unnumrem}
If we are interested in real Lax matrices we have to include also
the case where $K_3(\alpha_{1},\alpha_2)=\begin{pmatrix}
\alpha_{1} & - \alpha_2 \\
\alpha_2 & \ \alpha_1 %
\end{pmatrix}$ and the corresponding YB map of proposition
\ref{sympYB}.
\end{unnumrem}

%%%%%%%%%%%%%%%%%%%%%%%%%%%%%%%%%%%%%%%%%%%%%%%%%%%%%%%%%%%%%%%%%%%%%%%%%%%%%%%%%%%%%%%%%
%%%%%%%%%%%%%%%%%%%%%%%%%%%%%%%%%%%%%%%%%%%%%%%%%%%%%%%%%%%%%%%%%%%%%%%%%%%%%%%%%%%%%%%%%
%%%%%%%%%%%%%%%%%%%%%%%%%%%%%%%%%%%%%%%%%%%%%%%%%%%%%%%%%%%%%%%%%%%%%%%%%%%%%%%%%%%%%%%%%
\subsection{Degenerate YB maps} \label{DGNYB}
Degenerate YB maps can arise when $K(\alpha)$ is not invertible. 
A way of constructing degenerate YB maps as limits of the
non-degenerate ones was presented in \cite{kp} for $K(\alpha)
=K(\beta)=Constant$. We will apply this method here as well for
$K(\alpha) \neq K(\beta)$.

We consider a function $K:V\rightarrow GL_2(\mathbb{C})$, $V \subset
\mathbb{C}^4$, depending from a parameter $\varepsilon$, such that
$K(\alpha,\varepsilon
)K(\beta,\varepsilon)=K(\beta,\varepsilon)K(\alpha,\varepsilon)$ and
$\underset{\varepsilon \rightarrow 0}{\lim}
\det{K(\alpha,\varepsilon )}=0$ for every $\alpha, \beta \in
\mathbb{C}^m$, $m \leq 4$. We construct the corresponding
non-degenerate YB map $R _{\bar{\alpha},\bar{\beta}}(\varepsilon )$
of proposition \ref{sympYB}. The limit of $R
_{\bar{\alpha},\bar{\beta}}(\varepsilon )$, for $\varepsilon
\rightarrow 0$, can lead to a rational degenerate YB map on
$\mathbb{C}^2\times \mathbb{C}^2$. The induced Poisson structure is
defined by the limit of the Sklyanin bracket. We apply this
construction in the next concrete example.

\subsubsection*{A generalization of the Adler-Yamilov map}

We consider the function $K:\mathbb{C}\rightarrow GL_2(\mathbb{C})$
with $K(\alpha_1)=K_{\alpha_1}=\begin{pmatrix}
\alpha_{1} & 0 \\
0 & \varepsilon %
\end{pmatrix}.$
The Casimir functions on $\mathcal{L}^2_{K(\alpha_1)}$ are :
$$f_0(X;K(\alpha_1))=x_{11}x_{22}-x_{12}x_{21}, \ \ f_1(X;K(\alpha_1))=\varepsilon  x_{11} + \alpha_1 x_{22}.$$
(Here we denote by $x_{ij}$ the elements of the matrix $X$). If we
set $f_0(X;K(\alpha_1))=\alpha_2$, $f_1(X;K(\alpha_1))=\alpha_3$ and
solve with respect to $x_{11}, \ x_{22}$ we have
\begin{equation*}
x_{11}=\frac{1}{2\varepsilon }(\alpha _{3}-(\alpha_{3}^{2}-
4\alpha_{1}\varepsilon (\alpha _{2}+x_{12}x_{21}))^{1/2}), \
x_{22}=\frac{1}{2\alpha _{1}}(\alpha _{3}+ (\alpha_{3}^{2}-4\alpha
_{1}\varepsilon (\alpha_{2}+ x_{12}x_{21}))^{1/2}).
\end{equation*}
By substituting this values to $X-\zeta K(\alpha_1)$ and renaming 
$x_{12}$, $x_{21}$ as $x_1$ and $x_2$ respectively, we obtain the
three-parametric Lax matrix
\begin{equation}
L(x_{1},x_{2};\bar{\alpha})=%
\begin{pmatrix}
\frac{\alpha _{3}-(\alpha_{3}^{2}-4\alpha _{1}\varepsilon (\alpha
_{2}+x_{1}x_{2}))^{1/2}}{2\varepsilon }-\alpha _{1}\zeta  & x_{1} \\
x_{2} & \frac{\alpha _{3}+(\alpha_{3}^{2}-4\alpha _{1}\varepsilon (\alpha
_{2}+x_{1}x_{2}))^{1/2}}{2\alpha _{1}}-\varepsilon \zeta
\end{pmatrix}
\label{laxadl}
\end{equation}
with $\bar{\alpha}=(\alpha_1,\alpha_2,\alpha_3)$, of the
non-degenerate YB map of proposition \ref{sympYB}
\begin{equation} \label{adlYB}
R_{\mathcal{\bar{\alpha},\bar{\beta}}}((x_1,x_2),(y_1,y_2))
=((u_1,u_2),(v_1,v_2)).
\end{equation}
Here $u_1, \ u_2, \ v_1, \ v_2$ are the corresponding elements
$u_{12}, \ u_{21}, \ v_{12}, \ v_{21}$ of the matrices:
\begin{eqnarray*}
[ u_{ij} ]:= ~ U &=& (\alpha_1\varepsilon YX-
\alpha_2K_{\alpha_1}K_{\beta_1}) ((\alpha_1\varepsilon
(YK_{\alpha_1}+K_{\beta_1}X)-\alpha_3
K_{\alpha_1}K_{\beta_1})^{-1}K_{\alpha_1} \\
\text{[} v_{ij}  ]: = ~ V &=&
K_{\alpha_1}^{-1}(YK_{\alpha_1}+K_{\beta_1}X-UK_{\beta_1}),
\end{eqnarray*}
for $X=L'_{K(\alpha_1)}(x_{1},x_{2};\bar{\alpha})\equiv
L(x_1,x_2;\bar{\alpha})+ \zeta K_{\alpha}$ and
$Y=L'_{K(\alpha_1)}(y_{1},y_{2};\bar{\beta})\equiv
L(y_1,y_2;\bar{\beta})+ \zeta K_{\beta}$.  
 
The limit of (\ref{adlYB}), for $\varepsilon \rightarrow 0$, gives the
degenerate 6-parametric Yang-Baxter map 
\linebreak
$\tilde
R_{{\bar{\alpha},\bar{\beta}}}((x_1,x_2),(y_1,y_2))=((\bar u
_1,\bar u _2),( \bar v _1, \bar v _2)), $
where  
\begin{eqnarray*}
 \bar{u}_{1} =\frac{\beta _{1}}{\alpha _{1}\beta _{3}}(\alpha _{3}y_{1}-Q x_1%
),\ \ \ \bar{u}_{2}=\frac{\alpha _{1}}{\beta _{1}}~y_{2}, \ \ \ 
\bar{v}_{1} =\frac{\beta _{1}}{\alpha _{1}}~x_{1}, \ \ \ 
\bar{v}_{2}=\frac{\alpha _{1}}{\beta _{1}\alpha _{3}}(\beta _{3}x_{2}-Q y_2), \ \ \ \\ 
\text{and} \ Q=\frac{\alpha _{1}\beta _{1}(\alpha _{2}\beta_{3}-\alpha _{3}\beta _{2})%
}{\alpha _{3}\beta _{3}+\alpha _{1}\beta _{1}x_{1}y_{2}} \ . \ \ \ \ \ \ \ \ \ \ \ \ \ \ \
 \ \ \ \ \ \ \ \ \ \ \ \ \ \ \ \ \ \ \ \ \ \  \ \ \ \ \ \ \ \ \ \ \ \ \ \ \
 \ \ \ \ \ \ \ \ \ \ \ \ \ \ \ \ \ \ \ \ \ \  \ \ \ 
\end{eqnarray*}
This map is symplectic with respect to the symplectic form obtained by taking
the limit, for $\varepsilon \rightarrow 0$, of $J_{K_{\alpha}}(L'(x_{1},x_{2};\bar{\alpha}))$
and $J_{K_{\alpha}}(L'(y_{1},y_{2};\bar{\beta}))$,
\begin{equation} \label{adp}
\{x_1,x_2\}=\alpha_3, \  \{y_1,y_2\}=\beta_3, \  \{x_i,y_j\}=0,
\end{equation}
and admits the strong Lax matrix
\begin{equation*}
M(x_{1},x_{2};\bar{\alpha})=\lim_{\varepsilon \rightarrow 0}L(x_{1},x_{2};%
\bar{\alpha})=%
\begin{pmatrix}
\frac{\alpha _{1}}{\alpha _{3}}(\alpha _{2}+x_{1}x_{2})-\alpha _{1}\zeta  &
x_{1} \\
x_{2} & \frac{\alpha _{3}}{\alpha _{1}}%
\end{pmatrix}.%
\end{equation*}

If we set $\alpha_3=\beta_3=1$ on the map $\tilde R_{{\bar{\alpha},\bar{\beta}}}$
we derive the 4-parametric YB map $\tilde R_{(\alpha_1,\alpha_2),(\beta_1,\beta_2)}$
with strong Lax matrix $M(x_{1},x_{2};\alpha_1,\alpha_2,1)$. The induced symplectic form
in this case is the canonical one.
Moreover by setting $\alpha_1=\beta_1=\alpha_3=\beta_3=1$, $\bar R_{{\bar{\alpha},\bar{\beta}}}$
is reduced to the Adler-Yamilov map \cite{adler,kp}.

According to \cite{pnc,kp2} the monodromy matrix of the
1-periodic `staircase' initial value problem on a quadrilateral
lattice is $M_1(x_1,x_2,y_1,y_2)\equiv
M(y_{1},y_{2};\bar{\beta})M(x_{1},x_{2};\bar{\alpha})$. The trace of
the monodromy matrix gives the two functionally independent
integrals :
\begin{eqnarray*}
J_1(x_1,x_2,y_1,y_2) &=& \frac{{\alpha_1}   {\beta_1}}{ \alpha_3}
{x_1} {x_2}+
\frac{  {\alpha_1}   {\beta_1} }{  \beta_3}  {y_1}   {y_2} \\
J_2(x_1,x_2,y_1,y_2) &=&   {x_2}   {y_1}+  {x_1}   {y_2}+ \frac{
{\alpha_1}   {\beta_1}} { {\alpha_3}
  {\beta_3}} (  {\alpha_2}+  {x_1}   {x_2}) (  {\beta_2}+  {y_1}   {y_2}).
\end{eqnarray*}
We can verify that these integrals are in involution with respect to
(\ref{adp}). So we conclude that the map $\tilde
R_{{\bar{\alpha},\bar{\beta}}}((x_1,x_2),(y_1,y_2)) \mapsto((\bar u
_1,\bar u _2),( \bar v _1, \bar v _2))$ is integrable in the
Liouville sense. For the Adler-Yamilov map the corresponding
integrals are given by setting $\alpha_1=\beta_1=\alpha_3=\beta_3=1$
in $J_1$ and $J_2$.
%%%%%%%%%%%%%%%%%%%%%%%%%%%%%%%%%%%%%%%%%%%%%%%%%%%%%%%%%%%%%%%%%%%%%%%%%%%%%%%%%%%%%%%%%%%%%%%%%%%%%%%%%%%%%%%%
%%%%%%%%%%%%%%%%%%%%%%%%%%%%%%%%%%%%%%%%%%%%%%%%%%%%%%%%%%%%%%%%%%%%%%%%%%%%%%%%%%%%%%%%%%%%%%%%%%%%%%%%%%%%%%%%
%%%%%%%%%%%%%%%%%%%%%%%%%%%%%%%%%%%%%%%%%%%%%%%%%%%%%%%%%%%%%%%%%%%%%%%%%%%%%%%%%%%%%%%%%%%%%%%%%%%%%%%%%%%%%%%%

\section{ Higher dimensional Yang-Baxter maps} \label{ndim}

In order to generate higher dimensional Yang-Baxter maps we consider
the set $\mathcal{L}^n$ of $n$ order polynomial matrices of the form
$X-\zeta A$. There are $n(n+1)$ functionally independent Casimir
functions on $\mathcal{L}^n$ with respect to the Sklyanin bracket
(\ref{sklyanin}), which are again the $n^2$ elements of $A$ and the
$n$ functions $f_{i},~i=0,...,n-1$, defined as the coefficients of
the polynomial $p_{X}^A(\zeta)=det(X-\zeta A)$,
\begin{equation*}\label{det}
p_{X}^A(\zeta)=(-1)^{n}f_{n}(X;A)\zeta^{n}+(-1)^{n-1}f_{n-1}(X;A)\zeta^{n-1}
+...+(-1)f_{1}(X;A)\zeta+f_{0}(X;A)
\end{equation*}
where $f_{n}(X;A)=detA$ and $f_{0}(X;A)=detX$.

As in the $2 \times 2$ case, we consider $K:\mathbb{C}^{d} \rightarrow
GL_n(\mathbb{C})$ a $d$--parametric family of commuting matrices.
Next, for $\alpha \in \mathbb{C}^{d}$, we denote the value $K(\alpha)$ by $K_{\alpha}$ and the
values of the Casimirs $f_{i}(X;K(\alpha))$ by $f_{i}(X;\alpha)$, $i=0,...,n$.
%%%%%%%%%%%%%%%%%%%%%%%%%%%%%%%%%%%%%%%%%%%%%%%%%%%%%%%%
%%%%%%%%%%%%%%%%%%%%%%%%%%%%%%%%%%%%%%%%%%%%%%%%%%%%%%%
\begin{prop} \label{nxn}
Let $U$ and $V$ be $n \times n$ matrices that satisfy the following two
conditions
\begin{itemize}
\item[\rm(i)] $f_{i}(U;\alpha)=f_{i}(X;\alpha)$ and  $
f_{i}(V;\beta)=f_{i}(Y;\beta)$ for $i=0,...,n-1$,
\item[\rm(ii)] $(U-\zeta K_{\alpha})(V-\zeta K_{\beta})=(Y-\zeta K_{\beta})(X-\zeta
 K_{\alpha})$, identically in $\zeta \in \mathbb{C}$
\end{itemize}
for $X, \ Y \in Mat(n \times n)$ such that $ \det
\sum_{i=1}^{n}(-1)^{i}f_{i}(X;\alpha)M_{i-1}
 \neq 0 \}$ 
. Then
 \begin{eqnarray} \label{Un}
U &=& \left(
-f_{0}(X;\alpha)I-\sum_{i=1}^{n}(-1)^{i}f_{i}(X;\alpha)N_{i-1}
\right)
\left(\sum_{i=1}^{n}(-1)^{i}f_{i}(X;\alpha)M_{i-1}\right) ^{-1}K_{\alpha} \\
V &=& K_{\alpha}^{-1}(YK_{\alpha}+K_{\beta}X-U K_{\beta}),
\label{Vn}
\end{eqnarray}
where $M_i$, $N_i$ are given by:
\begin{eqnarray*}
M_{0} &=& I, \ N_{0}=0, \
M_{1}=(YK_{\alpha}+K_{\beta}X)K_{\beta}^{-1}K_{\alpha}^{-1}, \  \
N_1=-YXK_{\beta}^{-1} K_{\alpha}^{-1},  \\
M_{i} &=& M_{1}M_{i-1}+N_{i-1}, \  \ N_{i}=N_{1}M_{i-1}, \ for \
i=2,...,n.
\end{eqnarray*}
\end{prop}

\bigskip

\paragraph{Proof:}
Since $f_{i}(U;\alpha)=f_{i}(X;\alpha)$, for $i=1,...,n$, then
$p_{U}^{K_\alpha}(\zeta)=p_{X}^{K_\alpha}(\zeta)$. Cayley-Hamilton
theorem states that
$p_{U}^{K_\alpha}(UK_{\alpha}^{-1})=p_{X}^{K_\alpha}(UK_{\alpha}^{-1})=0$.
So
\begin{equation}
\sum_{i=1}^{n}(-1)^{i}f_{i}(X;\alpha)(UK_{\alpha}^{-1})^{i}=
-f_{0}(X;\alpha)I,\ \ \ \ \ \ \ i=1,...,n. \label{caley}
\end{equation}
Furthermore from $(ii)$ we derive the system:
\begin{equation} \label{system}
UV=YX, \ \  UK_{\beta}+K_{\alpha}V=YK_{\alpha}+K_{\beta}X
\end{equation}
which implies
\begin{equation} \label{tetr3}
(UK_{\alpha}^{-1})^2
=UK_{\alpha}^{-1}(YK_{\alpha}+K_{\beta}X)K_{\beta}^{-1}K_{\alpha}^{-1}
-YXK_{\beta}^{-1} K_{\alpha}^{-1}.
\end{equation}
For simplicity we set $\tilde{U}=UK_{\alpha}^{-1},\
M_{1}=(YK_{\alpha}+K_{\beta}X)K_{\beta}^{-1}K_{\alpha}^{-1}$ and
$N_1=-YXK_{\beta}^{-1} K_{\alpha}^{-1}$.
So equation (\ref{tetr3}) can be written as $\tilde{U}^{2} =
\tilde{U}M_{1}+N_{1}$.
Also if we set $M_{0}=I,  N_{0}=0$ and define $M_{i}, N_{i}$ from
the recurrence relations:
\begin{eqnarray}
M_{i}=M_{1}M_{i-1}+N_{i-1},\ \ N_{i}=N_{1}M_{i-1} \ \ \ for \ \
i=1,...,n,\label{recurs}
\end{eqnarray}
then we can evaluate the powers of $\tilde{U}^{k}$ as $\tilde{U}^{k}
= \tilde{U}M_{k-1}+N_{k-1}$ for $k=1,...,n$.
So equation (\ref{caley}) becomes:
\begin{equation*}
\sum_{i=1}^{n}(-1)^{i}f_{i}(X;\alpha)(\tilde{U}M_{i-1}+N_{i-1})=
-f_{0}(X;\alpha)I,
\end{equation*}
and finally we have
\begin{equation*}
\tilde{U}=\left(
-f_{0}(X;\alpha)I-\sum_{i=1}^{n}(-1)^{i}f_{i}(X;\alpha)N_{i-1}
\right) \left(\sum_{i=1}^{n}(-1)^{i}f_{i}(X;\alpha)M_{i-1}\right)
^{-1}
\end{equation*}
So $U=\tilde{U}K_{\alpha}$ and from(\ref{system}) $V =
K_{\alpha}^{-1}(YK_{\alpha}+K_{\beta}X-U K_{\beta}).$

\begin{rem} \label{rem41}
If we write the first equation of (\ref{system}) as
$UK_{\alpha}^{-1}K_{\alpha}V=YX$ and replace $K_{\alpha}V$ from the
second one, we get that
$$ UK_{\alpha}^{-1}(YK_{\alpha}-UK_{\beta})=(YK_{\alpha}-UK_{\beta})K_{\alpha}^{-1}X.$$
In a similar way we can show that
$(UK_{\beta}-YK_{\alpha})K_{\beta}^{-1}V=YK_{\beta}^{-1}(UK_{\beta}-YK_{\alpha})$.
So if $\det(UK_{\beta}-YK_{\alpha}) \neq 0$ (equivalently
$\det(K_{\alpha}V-K_{\beta}X) \neq 0$ since $UK_{\beta}-YK_{\alpha}=
K_{\alpha}V-K_{\beta}X$) then the matrices $ UK_{\alpha}^{-1}, \
K_{\beta}^{-1}V$ are similar with the matrices $K_{\alpha}^{-1}X$
and $YK_{\beta}^{-1}$ respectively, and subsequently
$p_{U}^{K_{\alpha}}(\zeta)=p_{X}^{K_{\alpha}}(\zeta), \
p_{V}^{K_{\beta}}(\zeta)=p_{Y}^{K_{\beta}}(\zeta)$. Therefore the
condition $(i)$ of proposition \ref{nxn} can be replaced by the
assumption $\det(UK_{\beta}-YK_{\alpha}) \neq 0$ (equivalently
$\det(K_{\alpha}V-K_{\beta}X) \neq 0$).
\end{rem}
\begin{rem} \label{rem42}
Proposition \ref{nxn} holds also if we replace $K_{\alpha}, \
K_{\beta}$ by two invertible matrices $A$ and $B$ respectively such
that $AB=BA$. The reason for restricting to the function $K$ is that
we are interested to consider $L(X;\alpha)=X-\zeta K_{\alpha}$ as a
Lax matrix of a YB map, otherwise we would have a Lax pair
$L(X;A)=X-\zeta A, \ M(Y;B)=Y-\zeta B$ with $L \neq M$ as in
\cite{kp2}.
\end{rem}

The Yang-Baxter property of this re-factorization solution, i.e. of
the map $$\mathcal{R}_{\alpha,\beta}(X,Y)\mapsto(U,V),$$ with $U, \
V$ defined by (\ref{Un}) and (\ref{Vn}), is still an open problem.
In low dimensions, for certain choices of the function $K$, this can
be checked by direct computation or by proposition \ref{anagea}. We
conjecture that this is true for any dimension. Anyway, since
$f_{i}(U;\alpha)=f_{i}(X;\alpha)$ and
$f_{i}(V;\beta)=f_{i}(Y;\beta)$, the map
$\mathcal{R}_{\alpha,\beta}$ can be reduced, as in $2\times 2$ case,
to a  map on $\mathbb{C}^{n(n-1)}\times \mathbb{C}^{n(n-1)}$ by the
restriction to the corresponding level sets of the $n$ Casimir
functions $f_i, \ i=0,...,n-1$. Further reduction on lower
dimensional symplectic leaves is also possible.
%-----------------------------------------------------------------------------------
%-----------------------------------------------------------------------------------
\subsection{ 8-dimensional quadrirational symplectic YB maps with $3\times 3$ Lax matrices}

In the case of $\mathcal{L}^3$ there exist three Casimir functions,
so the map of Prop.\ref{nxn} can be reduced to a  quadrirational map
on $\mathbb{C}^6 \times \mathbb{C}^6$. Further reduction to four
dimensional symplectic submanifolds of $\mathcal{L}^3$ provide maps
on $\mathbb{C}^4 \times \mathbb{C}^4$. Next, we demonstrate this
procedure for $K_{\alpha}=K_{\beta}=I$.
%%%%%%%%%%%%%%%%%%%%%%%%%%%%%%%%%%%
%%%%%%%%%%%%%%%%%%%%%%%allagh-X-A
Let $L(\zeta )=X- \zeta I$, with $X=[x_{ij}]$, be a generic
%allagh-X-A
element of $\mathcal{L}^3 _{I}$.
In this case the Sklyanin bracket is
\begin{eqnarray}
 \{L(\zeta ) \ \overset{\otimes }{,} \ L(\eta)\}=[\frac{r}{\zeta
-\eta},L(\zeta )\otimes L(\eta)] \nonumber
\ \ \ \ \ \ \ \ \ \ \ \ \ \ \ \ \ \ \ \ \ \ \ \ \ \ \ \ \\ \nonumber \\
=
{
%\scriptsize 
 \left(
\begin{array}{ccccccccc}
 0 & -x_{12} & -x_{13} & x_{12} & 0 & 0 & x_{13} & 0 & 0 \\
 x_{21} & 0 & 0 & x_{22}-x_{11} & -x_{12} & -x_{13} & x_{23} & 0 & 0 \\
 x_{31} & 0 & 0 & x_{32} & 0 & 0 & x_{33}-x_{11} & -x_{12} & -x_{13} \\
 -x_{21} & x_{11}-x_{22} & -x_{23} & 0 & x_{12} & 0 & 0 & x_{13} & 0 \\
 0 & x_{21} & 0 & -x_{21} & 0 & -x_{23} & 0 & x_{23} & 0 \\
 0 & x_{31} & 0 & 0 & x_{32} & 0 & -x_{21} & x_{33}-x_{22} & -x_{23} \\
 -x_{31}& -x_{32} & x_{11}-x_{33} & 0 & 0 & x_{12} & 0 & 0 & x_{13} \\
 0 & 0 & x_{21} & -x_{31} & -x_{32} & x_{22}-x_{33} & 0 & 0 & x_{23} \\
 0 & 0 & x_{31} & 0 & 0 & x_{32} & -x_{31} & -x_{32} & 0
\end{array}
\right)} \nonumber  \\  \label{skl3}
\end{eqnarray}

Generically the rank of the structure matrix (\ref{skl3})
is six. We are interested in finding
4-dimensional symplectic submanifolds of $\mathcal{L}^3 _{I}$.
For this reason we would like to
find conditions such that the rank of the matrix (\ref{skl3}) drops down to four.

Let $i_1<...<i_6, \ j_1<...<j_6$, with $i_k, \ j_k \in \{ 1,...,9 \}$ for
$k=1,...,6$. We denote by
$m((i_1,...,i_6), (j_1,...j_6) )$ the sixth order minor of the matrix (\ref{skl3}),
consisting of the $i_1,...,i_6$ rows and the
$j_1,...,j_6$ columns. Using this notation we prove the next lemma.
\begin{lem}
Consider the system of equations obtained by setting all sixth order minors  $m((i_1,...,i_6),
(j_1,...j_6) )$ equal to zero. There is a unique solution of this
system  with respect to $x_{11}, \ x_{31}, \ x_{32}$, for nonzero $x_{13}, \
x_{23}$, namely:
\begin{eqnarray} \label{minors}
x_{11} &=& \frac{ {x_{13}}  {x_{21}}}{ {x_{23}}}+ {x_{22}}-\frac{ {x_{12}}  {x_{23}}}{ {x_{13}}} \ , \
x_{31}=\frac{ {x_{21}} ( {x_{12}}  {x_{23}}+ {x_{13}} ( {x_{33}}- {x_{22}}))}{ {x_{13}}  {x_{23}}} \ , \nonumber \\
x_{32} &=& \frac{ {x_{12}} ( {x_{12}}  {x_{23}}+ {x_{13}} ( {x_{33}}- {x_{22}}))}{ {x_{13}}^2} \ .
\end{eqnarray}
Substituting these values to $X-\zeta I$ the rank of the Poisson
matrix in (\ref{skl3}) reduces to four and the Casimirs
$f_0(X;I):=\alpha_0, \ f_1(X;I):=\alpha_1$,  $f_2(X;I):=\alpha_2$
satisfy
\begin{equation} \label{surf}
 4 {\alpha_0} {\alpha_2}^3-{\alpha_1}^2 {\alpha_2}^2+4 {\alpha_1}^3-18
{\alpha_0} {\alpha_1} {\alpha_2}+27 {\alpha_0}^2=0
\end{equation}
\end{lem}

\paragraph{Proof:}
Consider the minors
\begin{eqnarray*}
m_1=m((1,2,3,4,5,6),(3,4,6,7,8,9))&=&-\left(x_{21} x_{13}^2-x_{11}
x_{23} x_{13}+x_{22} x_{23} x_{13}-x_{12} x_{23}^2\right)^2, \\
m_2=m((1,2,3,4,6,7),(3,4,5,6,8,9))&=&-\left(x_{23} x_{12}^2-x_{13}
x_{22} x_{12}+x_{13} x_{33} x_{12}-x_{13}^2 x_{32}\right)^2, \\
m_3=m((1,2,3,5,6,9),(1,2,3,5,6,9))&=&-(x_{12} x_{23} x_{31}-
x_{13} x_{21} x_{32})^2.
\end{eqnarray*}
The system $m_1=m_2=m_3=0$ is linear with respect to $x_{11}, \ x_{31}, \ x_{32}$
and for $x_{13}, \ x_{23} \neq 0$ admits the unique solution (\ref{minors}).
Substituting  these values to (\ref{skl3}) the rank reduces to four and
the Casimir functions become:
\begin{eqnarray} \label{cas3}
f_0(X;I) &=& \frac{(x_{13} x_{22}-x_{12} x_{23})^2 \left(x_{21} x_{13}^2+x_{23} x_{33}
   x_{13}+x_{12} x_{23}^2\right)}{x_{13}^3 x_{23}} \nonumber \\
f_1(X;I) &=& \frac{(x_{13} x_{22}-x_{12} x_{23}) \left(2 x_{21} x_{13}^2+x_{23} (x_{22}+2 x_{33})
   x_{13}+x_{12} x_{23}^2\right)}{x_{13}^2 x_{23}} \\
f_2(X;I) &=& \frac{x_{13} x_{21}}{x_{23}}+2 x_{22}-\frac{x_{12} x_{23}}{x_{13}}+x_{33}. \nonumber
\end{eqnarray}
which satisfy (\ref{surf}).

%_____________________________________________________________
\begin{figure}
%\centering
\includegraphics[scale=0.89]{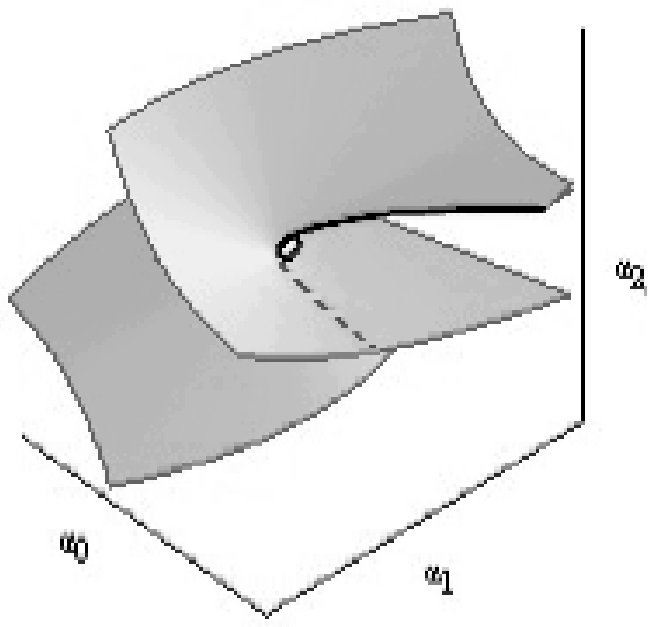}
\includegraphics[scale=0.89]{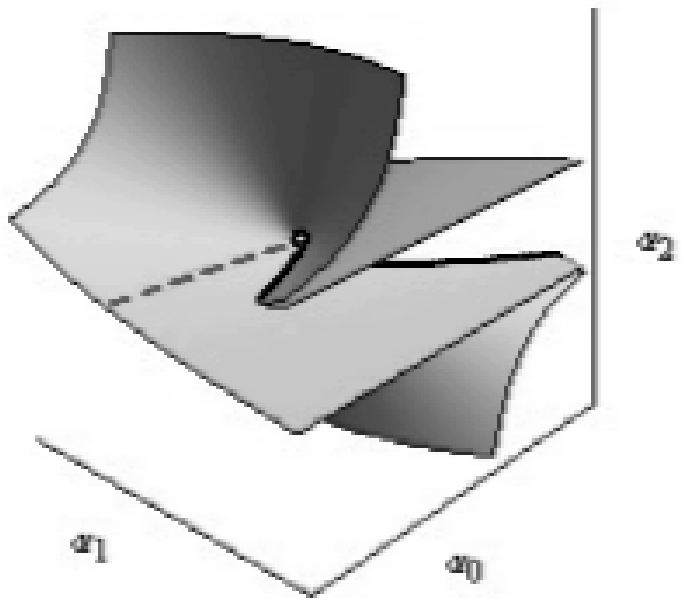}
\caption{ Two views of surface (\ref{surf}) in $\mathbb{R}^3$, black curve:
$(\alpha^3,3\alpha^2,3\alpha)$, dashed curve: $(-\alpha^3,
-\alpha^2,\alpha)$ } \label{surfcas}
\end{figure}
%_____________________________________________________________

It is remarkable that two curves on the surface (\ref{surf}) give rise to maps related to the
Boussinesq and the matrix KdV equation.

%%%%%%%%%%%%%%%%%%%%%%%%%%%%%%%%%%%%%%%%%%%%%%%%%%%%%%%%%%%%%%%%%%%%%%%%%%%%%%%%%%%%%%%%
%%%%%%%%%%%%%%%%%%%%%%%%%%%%%%%%%%%%%%%%%%%%%%%%%%%%%%%%%%%%%%%%%%%%%%%%%%%%%%%%%%%%%%%
\subsubsection{A 4-parametric symplectic Y-B map}
If we set the values (\ref{minors}) to $X$, in order to restrict on
the level sets of the Casimir functions of $\mathcal{L}^3 _{I}$ we
set $f_2(X;I)=\alpha_2, \ f_1(X;I)=\alpha_1$ (of course $f_0(X;I)$
will be also constant since (\ref{surf}) must be satisfied) and
solve (\ref{cas3}) with respect to $x_{22}$ and $x_{33}$ to get
\begin{eqnarray*}
x_{22} =\frac{\alpha_2}{3}+\frac{x_{12}x_{23}}{x_{13}}\pm \frac{1}{3}\sqrt{\alpha_2^{2}-3\alpha_1} \ , \
x_{33} =\frac{\alpha_2}{3}-\frac{x_{13}x_{21}}{x_{23}}-\frac{x_{12}x_{23}}{x_{13}}\mp
\frac{2}{3}\sqrt{\alpha_2^{2}-3\alpha_1} \ .
\end{eqnarray*}
For simplicity we can change the parameters into $c_1=\frac{\alpha_2}{3}$ and
$c_2=\pm \frac{1}{3} \sqrt{\alpha_2^{2}-3\alpha_1}$, so
$x_{22} =c_1+c_2+\frac{x_{12}x_{23}}{x_{13}} , \
x_{33} =c_1-2c_2-\frac{x_{13}x_{21}}{x_{23}}-\frac{x_{12}x_{23}}{x_{13}}$.
Substituting these values to (\ref{minors}) and the new $x_{ij}$ to $X-\zeta I$,
we obtain the two parametric family of matrices

\bigskip

$M(x_{12} , x_{13},x_{21},x_{23};c_1,c_2)$
\begin{equation}
=\begin{pmatrix}
 \frac{x_{13} x_{21}}{x_{23}}+c_{1}+c_{2}-\zeta  & x_{12} & x_{13}
   \\
 x_{21} & \frac{x_{12} x_{23}}{x_{13}}+c_{1}+c_{2}-\zeta  & x_{23}
   \\
 -\frac{x_{13} x_{21}^2}{x_{23}^2}-\frac{3 c_{2}
   x_{21}}{x_{23}}-\frac{x_{12} x_{21}}{x_{13}}
& -\frac{x_{23} x_{12}^2}{x_{13}^2}-\frac{3 c_{2}
   x_{12}}{x_{13}}-\frac{x_{21} x_{12}}{x_{23}}
&c_{1}-2 c_{2} -\frac{x_{13} x_{21}}{x_{23}}-\frac{x_{12}
   x_{23}}{x_{13}}-\zeta
\end{pmatrix}
\end{equation}
The reduced Poisson structure is
\begin{equation*}
\{x_{12},x_{21} \}=\frac{x_{12}x_{23}}{x_{13}}-\frac{x_{13}x_{21}}{x_{23}}, \
\{x_{12},x_{23} \}=-x_{13}, \ \{x_{13},x_{21} \}=x_{23}
\end{equation*}
and $\{x_{12},x_{13}\}=\{x_{13},x_{23}\}=\{x_{21},x_{23}\}=0$, which defines the symplectic form :
\begin{equation*}
\omega =\frac{1}{x_{23}}dx_{13}\wedge dx_{21}-\frac{1}{x_{13}}dx_{12}\wedge
dx_{23}+(\frac{x_{12}}{x_{13}^{2}}-\frac{x_{21}}{x_{23}^{2}})dx_{13}\wedge
dx_{23}\ .
\end{equation*}
We can change to canonical variables by setting
\begin{equation} \label{trns}
x_{13}=X_1, \ x_{23}=X_2, \ x_{21}=-x_1 X_2 , \ x_{12}=-x_2 X_1.
\end{equation}
Then we denote matrix $M(x_{12} , x_{13},x_{21},x_{23};c_1,c_2)$ by \\
%%%%%%%%%%%%%%%%%ALLAGH-STOP 1373
$L(x_1,x_2,X_1,X_2;c_1,c_2) \equiv L'_I(x_1,x_2,X_1,X_2;\alpha_1,\alpha_2) - \zeta I $
\begin{eqnarray}
\equiv \begin{pmatrix}
 c_{1}+c_{2}-x_{1} X_{1} - \zeta & -X_{1} x_{2} & X_{1} \\
 -x_{1} X_{2} & c_{1}+c_{2}-x_{2} X_{2} -\zeta & X_{2} \\
 -x_{1} (x_{1} X_{1}+x_{2} X_{2}-3 c_{2}) & -x_{2} (x_{1} X_{1}+x_{2} X_{2}-3 c_{2}) &
   c_{1}-2 c_{2}+x_{1} X_{1}+x_{2} X_{2} -\zeta
\end{pmatrix}
  \label{bougon}
\end{eqnarray}
and the symplectic form $\omega$ by the canonical symplectic form
$\omega_0= dx_1\wedge dX_1+dx_2\wedge dX_2.$

From the
re-factorization formula (\ref{Un}), (\ref{Vn}), for
$K_{\alpha}=K_{\beta}=I$,  $X =
%$ K_{\alpha}=K_{\beta}=I$, \ $X =
L'_I(x_1,x_2,X_1,X_2;\alpha_1,\alpha_2) $ and  $Y =
L'_I(y_1,y_2,Y_1,Y_2;\beta_1,\beta_2 ) $, 
since the Casimir functions on
$$\Sigma_I(\alpha_1,\alpha_2)=\{L(x_1,x_2,X_1,X_2;\alpha_1,\alpha_2)
\ | \ x_1,x_2,X_1,X_2 \in \mathbb{C}\}$$ are
$$f_0(X;I)=(\alpha_1-2\alpha_2)(a_1+a_2)^2, \ f_1(X;I)=3(\alpha_1^2-\alpha_2^2), \ f_2(X;I)=3\alpha_1,$$
 we obtain the matrices

$$U=(YX(3\alpha_1 I
-Y-X)- (\alpha_1-2\alpha_2)(a_1+a_2)^2 I) ((3\alpha _1
I-Y-X)(Y+X)+YX-3(\alpha_1^2-\alpha_2^2) I)^{-1},$$ $V=Y+X-U$.
\\
If we denote by $U_{ij}, V_{ij}$ the elements of the matrices $U$
and $V$, we come up to the next proposition.
\begin{prop} \label{gen3}
The map

$R_{((\alpha_1,\alpha_2),(\beta_1,\beta_2))}:((x_1,x_2,X_1,X_2),(y_1,y_2,Y_1,Y_2))
\mapsto ((u_1,u_2,U_1,U_2),(v_1,v_2,V_1,V_2))$
where
\begin{eqnarray*}
U_1 &=& U_{13}, \ U_2 = U_{23}, \ u_1=-\frac{U_{21}}{U_{23}}, \ u_2=-\frac{U_{12}}{U_{13}}, \\
V_1 &=& V_{13},\ V_2 = V_{23}, \ \ v_1=-\frac{V_{21}}{V_{23}}, \ v_2=-\frac{V_{12}}{V_{13}}
\end{eqnarray*}
is a symplectic parametric Yang-Baxter map, with respect to the
canonical symplectic form $dx_1\wedge dX_1+dx_2\wedge
dX_2+dy_1\wedge dY_1+dy_2\wedge dY_2$, and admits the strong Lax
matrix $L(x_1,x_2,X_1,X_2;\alpha_1,\alpha_2)$.
\end{prop}

\paragraph{Proof:}
The YB property of this map can be checked by direct computation.
Moreover $u_i, \ U_i, \ v_i, \ V_i, \ i=1,2$ is the unique solution
(proposition \ref{nxn}) of the Lax equation:
$$L(u_1,u_2,U_1,U_2;\alpha_1,\alpha_2)L(v_1,v_2,V_1,V_2;\beta_1,\beta_2)=
L(y_1,y_2,Y_1,Y_2;\beta_1,\beta_2)L(x_1,x_2,X_1,X_2;\alpha_1,\alpha_2)$$

The explicit formula of the YB map
$R_{((\alpha_1,\alpha_2),(\beta_1,\beta_2))}$ of proposition
\ref{gen3} is
\begin{eqnarray*}
(u_1,u_2)&=&(y_1,y_2)-\frac{\alpha_1 -\beta_1-2 (\alpha_2  -
\beta_2)}{D} \ (x_1-y_1,x_2-y_2 ), \\
(v_1,v_2)&=&(x_1,x_2)+\frac{\alpha_1 -\beta_1+\alpha_2  -\beta_2}{D}
\  \ (x_1-y_1,x_2-y_2 ),
\end{eqnarray*}
with $D = 2 \alpha_2-\alpha_1+\beta_1+\beta_2+y_1 X_1+ y_2 X_2-x_1
X_1-x_2 X_2$, and
\begin{eqnarray*}
U_1=\frac{ (x_1-v_1)X_1+(y_1-v_1)Y_1}{ u_1 -v_1} , \ \
U_2=\frac{ (x_2-v_2)X_2+(y_2-v_2)Y_2}{ u_2 -v_2}   , \\
V_1=\frac{ (x_1-u_1)X_1+(y_1-u_1)Y_1}{ v_1 -u_1}  , \ \
V_2=\frac{(x_2-u_2)X_2+(y_2-u_2)Y_2}{ v_2 -u_2}  .
\end{eqnarray*}

We will point out two special cases of this YB map that give rise
to Boussinesq and Goncharenko--Veselov maps.
%%%%%%%%%%%%%%%%%%%%%%%%%%%%%%%%%%%%%%%%%%%%%%%%%%%%%%%%%%%%%%%%%%%%%%%%%%%%%%%%%%%%%%%%
%%%%%%%%%%%%%%%%%%%%%%%%%%%%%%%%%%%%%%%%%%%%%%%%%%%%%%%%%%%%%%%%%%%%%%%%%%%%%%%%%%%%%%%%
\subsubsection{The Boussinesq Y-B map ($ \alpha_0=\alpha^3, \ \alpha_1=3\alpha^2, \ \alpha_2=3\alpha $)}
By setting $c_2=0, \ c_1=\alpha$ to (\ref{bougon}) we derive the Lax
matrix
\begin{equation*}
L_B(x_1,x_2,X_1,X_2;\alpha)=
\begin{pmatrix}
\alpha-\zeta -x_1 X_1 & -X_1 x_2 & X_1 \\
-x_1 X_2 & \alpha-\zeta -x_2 X_2 & X_2 \\
-x_1(x_1 X_1+x_2 X_2) & -x_2(x_1 X_1+x_2 X_2) & \alpha-\zeta +x_1
X_1+x_2 X_2
\end{pmatrix}
\end{equation*}
In this case the Casimir functions on
$\Sigma_I(\alpha)=\{L^B(x_1,x_2,X_1,X_2;\alpha) \ / \
x_1,x_2,X_1,X_2 \in \mathbb{C}\}$ are
$$f_0(X;I)=\alpha^3, \ f_1(X;I)=3\alpha^2,
\ f_2(X;I)=3\alpha,$$ for $X = L_{I}^B(x_1,x_2,X_1,X_2;\alpha)
\equiv L_B(x_1,x_2,X_1,X_2;\alpha)+ \zeta I$. The curve
$(\alpha^3,3\alpha^2,3\alpha)$ is depicted in fig.  \ref{surfcas}
with black color.

The corresponding 2-parametric YB map $R^B_{\alpha,\beta}$ with
strong Lax matrix $L_B(x_1,x_2,X_1,X_2;c)$ is induced from the YB
map $R_{((\alpha_1,\alpha_2),(\beta_1,\beta_2))}$ of proposition
\ref{nxn} i.e. $R^B_{\alpha,\beta}=R_{((\alpha,0),(\beta,0))}$ .

%--------------------------------------------------------------------------------------------------------------
%--------------------------------------------------------------------------------------------------------------
\subsubsection{ The Goncharenko--Veselov map ($\alpha_0=-\alpha^3, \ \alpha_1=-\alpha^2, \alpha_2=\alpha$)}
In a similar way if we set $c_1=\frac{\alpha}{3}$ and
$c_2=\frac{2\alpha}{3}$  we obtain the Yang-Baxter map
$$R^{GV}_{\alpha,\beta}=R_{((\frac{\alpha}{3},\frac{2\alpha}{3}),(\frac{\beta}{3},\frac{2\beta}{3}))}$$
with strong Lax matrix
%%%%%%%%%%%%%%%%%%%%%%%%%%%%%%%%%%%
\begin{equation*} \label{laxgon}
L_{GV}(\mathbf{x};\alpha)=
\begin{pmatrix}
\alpha-\zeta -x_1 X_1 & -X_1 x_2 & X_1 \\
-x_1 X_2 & \alpha-\zeta -x_2 X_2 & X_2 \\
-x_1(x_1 X_1+x_2 X_2-2\alpha) & -x_2(x_1 X_1+x_2 X_2-2\alpha) & x_1
X_1+x_2 X_2- \alpha-\zeta
\end{pmatrix}
\end{equation*}
%%%%%%%%%%%%%%%%%%%%%%%%%%%%%%%%%%%%%%%%%%%
for $\mathbf{x}=(x_1,x_2,X_1,X_2)$. Here for
$X=L_{GV}(\mathbf{x};\alpha)+ \zeta I$,
$(f_0(X;I),f_1(X;I),f_2(X;I))=(-\alpha^3,-\alpha^2,\alpha)$, which
is the dashed curve of fig. \ref{surfcas}.

Both maps $R^B_{\alpha,\beta}$ and $R^{GV}_{\alpha,\beta}$
are symplectic with respect to the canonical symplectic form $dx_1\wedge dX_1+dx_2\wedge dX_2+dy_1\wedge dY_1+dy_2\wedge dY_2$.

In \cite{gon}, Goncharenko and Veselov presented a YB map as interaction of two soliton solutions of
the matrix KdV equation and claimed that it admits the Lax matrix of the form:
\begin{equation} \label{laxgon2}
A(\xi ,\eta ; \lambda )=I+\frac{2 \lambda}{\zeta-\lambda} \frac{\xi\otimes \eta}{(\xi,\eta)} \ ,
\end{equation}
for the n-dimensional vectors $\xi$ and $\eta$. Here $\lambda$ is the YB parameter.
Essentially $\xi ,  \eta \in \mathbb{C}P^{n-1}$ since $\xi\mapsto \mu \xi$,
$\eta\mapsto \nu \eta$ leaves (\ref{laxgon2}) invariant.
Even if the case for $n=2$ is rather trivial, it is quite interesting for
higher dimensions.

First we observe that we can multiply the Lax matrix (\ref{laxgon2})
with ${\zeta-\lambda}$ and change $\zeta$ with $-\zeta$
in order to derive an equivalent Lax matrix
$$ B(\xi ,\eta; \lambda )=\lambda ( 2 \  \frac{ \xi\otimes \eta}{(\xi,\eta)}-I) - \zeta I $$
for the same YB map.
Now, let $n=3$, $\xi=(\xi_1,\xi_2,\xi_3)$ and
$\eta=(\eta_1,\eta_2,\eta_3)$.
Considering the affine part of $\mathbb{C}P^2$, we have $\xi=(\xi_1,\xi_2,1)$,
$\eta=(\eta_1,\eta_2,1)$ and by performing the invertible transformation
$(\eta_1,\eta_2,\xi_1,\xi_2,)\mapsto (x_1,x_2,X_1,X_2)$:
$$x_{1} =-\eta _{1} \ ,\
x_{2} =-\eta _{2} \ , \
X_{1}=\frac{2\alpha \xi _{1}} {\xi _{1}\eta _{1}+\xi _{2}\eta _{2}+1} \ ,
 \ X_{2}=\frac{2\alpha
\xi_{2}}{\xi _{1}\eta _{1}+\xi _{2} \eta _{2}+1}\ ,$$
the matrix $B(\xi ,\eta; \lambda )$ is transformed to the Lax matrix $L_{GV}(\mathbf{x};-\lambda)$.
%--------------------------------------------------------------------------------------------------------------
%--------------------------------------------------------------------------------------------------------------

%--------------------------------------------------------------------------------------------------------
\section{Conclusion} \label{persp}

By generalizing the re-factorization procedure reported in \cite{kp}, we presented a 
construction of multidimensional parametric Yang-Baxter maps. The symplectic quadrirational YB maps 
on $\mathbb{C}^2 \times \mathbb{C}^2$, that was derived in this way, where classified in two cases 
(three cases for real maps). 
The re-factorization of $3 \times 3$ binomial matrices provided us 
a family of symplectic YB maps on 
$\mathbb{C}^4 \times \mathbb{C}^4$ with Lax matrices the four dimensional symplectic leaves 
of $\mathcal{L}^3_I$.  
 
A similar classification procedure with the one presented here  
for quadrirational YB maps with $n \times n$ 
binomial Lax matrices, for $n>2$,  
is a far more difficult task. The determination of the commuting pairs of invertible $n \times n$ 
matrices, in addition with the determination of the corresponding symplectic leaves on $\mathcal{L}^n$, is needed. 
It would be interesting to investigate this problem for small values of $n$. 
Furthermore other re-factorization formulas of higher degree polynomial matrices, guided by the invariance of 
the Casimir functions of the Sklyanin bracket, could lead to symplectic multidimensional YB maps.
The derived maps contain, in general, more than one YB parameters. One can ask if (some of) these parameters are associated to spectral ones, in view of the 3D consistency of the YB maps. This is an interesting question especially with respect to finding invariants of the corresponding transfer maps and is going to be investigated in the future. Other issues deserving further research are  initial value problems on lattices connected to the maps reported here,
as well as the study of their continuum limits.

\paragraph{acknowledgments}
TEK acknowledges partial support from the State
Scholarships Foundation of Greece. Both authors thank the anonymous referee for useful comments.

%%%%%%%%%%%%%%%%%%%%%%%%%%%%%%%%%%%%%%%%%%%%%%%%%%%%%%%%%%%%%%%%%%%%%%%%%%%%%%%%%%%%%%%%%%%%%%%%%%%%%%%%%%%%%%%
%%%%%%%%%%%%%%%%%%%%%%%%%%%%%%%%%%%%%%%%%%%%%%%%%%%%%%%%%%%%%%%%%%%%%%%%%%%%%%%%%%%%%%%%%%%%%%%%%%%%%%%%%%%%%%

%--------------------------------------------------------------------------------------------------------------
%--------------------------------------------------------------------------------------------------------------

%------------------------------------------------------------------------------------------------------------------
\end{document}